\documentclass[10pt, conference, compsocconf]{IEEEtran}
\usepackage{graphicx}
\ifCLASSINFOpdf
\else
\fi
\usepackage[cmex10]{amsmath}
\begin{document}
%
\title{Geometry and Morphology of the Cosmic Web:\\
Analyzing Spatial Patterns in the Universe}


\author{\IEEEauthorblockN{Rien van de Weygaert, Bernard J.T. Jones, Erwin Platen}
\IEEEauthorblockA{Kapteyn Astronomical Institute\\
University of Groningen\\
P.O. Box 800, 9700 AV Groningen, the Netherlands\\
Email: weygaert@astro.rug.nl}
\and
\IEEEauthorblockN{Miguel A. Arag\'on-Calvo}
\IEEEauthorblockA{Dept. Physics \& Astronomy\\
The Johns Hopkins University\\
3701 San Martin Drive, Baltimore, MD 21218, USA\\
Email: miguel@pha.jhu.edu}
}

%

\IEEEspecialpapernotice{(Invited Paper)}

\maketitle

\begin{abstract}
We review the analysis of the Cosmic Web by means of an extensive toolset based on 
the use of Delaunay and Voronoi tessellations. The Cosmic Web is the salient 
and pervasive foamlike pattern in which matter has organized itself on 
scales of a few up to more than a hundred Megaparsec. The weblike spatial arrangement 
of galaxies and mass into elongated filaments, 
sheetlike walls and dense compact clusters, the existence of large near-empty void 
regions and the hierarchical nature of this mass distribution are three 
major characteristics of the comsic matter distribution. 

First, we describe the Delaunay Tessellation Field Estimator.Using the unique adaptive 
qualities of Voronoi and Delaunay tessellations, DTFE infers the density 
field from the (contiguous) Voronoi tessellation of a sampled galaxy 
or simulation particle distribution and uses the Delaunay tessellation as 
adaptive grid for defining continuous volume-filling fields of density and 
other measured quantities through linear interpolation. The resulting DTFE formalism 
is shown to recover the hierarchical nature and the anisotropic morphology 
of the cosmic matter distribution. The Multiscale Morphology Filter (MMF) uses the 
DTFE density field to extract the diverse morphological elements - filaments, 
sheets and clusters - on the basis of a ScaleSpace analysis which searches 
for these morphologies over a range of scales. Subsequently, we discuss the 
Watershed Voidfinder (WVF), which invokes the discrete watershed transform 
to identify voids in the cosmic matter distribution. The WVF is able to 
determine the location, size and shape of the voids. The watershed transform 
is also a key element in the SpineWeb analysis of the cosmic matter 
distribution. Finding its mathematical foundation in Morse theory, it allows 
the determination of the filamentary spine and connected walls in the 
cosmic matter density field through the identification of the 
singularities and corresponding separatrices. The first results of a direct 
implementation on the Delaunay tessellation itself are presented. Finally, 
we describe the concept of Alphashapes for assessing the topology of the 
cosmic matter distribution. 
\end{abstract}


%
\IEEEpeerreviewmaketitle

\section{Introduction: the Cosmic Web}
The large scale distribution of matter revealed by galaxy surveys features a complex 
network of interconnected filamentary galaxy associations. This network, which has become known as 
the {\it Cosmic Web} \cite{bondweb1996}, contains structures from a few megaparsecs\footnote{The main 
measure of length in astronomy is the parsec. Technically a parsec is the distance at which we would 
see the distance Earth-Sun at an angle of 1 arcsec. It is equal to 3.262 lightyears 
$=3.086\times 10^{13} \hbox{\rm km}$. Cosmological distances are substantially larger, so that a 
Megaparsec ($=10^6\,pc$) is the regular unit of distance. Usually this goes along with $h$, the 
cosmic expansion rate (Hubble parameter) $H$ in units of $100$ km/s/Mpc ($h\approx 0.71$).} up to tens 
and even hundreds of Megaparsecs of size. Galaxies and mass exist in a wispy weblike spatial arrangement 
consisting of dense compact clusters, elongated filaments, and sheetlike walls, amidst large near-empty 
void regions, with similar patterns existing at earlier epochs, albeit over smaller scales. The 
hierarchical nature of this mass distribution, marked by substructure over a wide range of scales and 
densities, has been clearly demonstrated. Its appearance has been most dramatically 
illustrated by the recently produced maps of the nearby cosmos, the 2dFGRS, the SDSS and the 2MASS redshift 
surveys \cite{colless2003,tegmark2004,huchra2005}\footnote{Because of the expansion of the Universe, 
any observed cosmic object will have its light shifted redward: its redshift $z$. According to Hubble's 
law, the redshift $z$ is directly proportional to the distance $r$ of the object, for $z\ll 1$: $cz=Hr$ 
(with $c$ the velocity of light, and $H\approx 71 km/s/Mpc$ the Hubble constant). Because it is extremely 
cumbersome to measure distances $r$ directly, cosmologists resort to the expansion of the Universe and 
use $z$ as a distance measure. Because of the vast 
distances in the Universe, and the finite velocity of light, the redshift $z$ of an object may also be 
seen as a measure of the time at which it emitted the observed radiation. }. 

The vast Megaparsec cosmic web is one of the most striking examples of complex geometric patterns 
found in nature, and certainly the largest in terms of sheer size. Computer simulations suggest that 
the observed cellular patterns are a prominent and natural aspect of cosmic structure formation through 
gravitational instability \cite{peebles80}, the standard paradigm for the emergence of structure in 
our Universe. 

\begin{figure*}
  \begin{center}
     \mbox{\hskip -0.5truecm\includegraphics[width=0.90\textwidth]{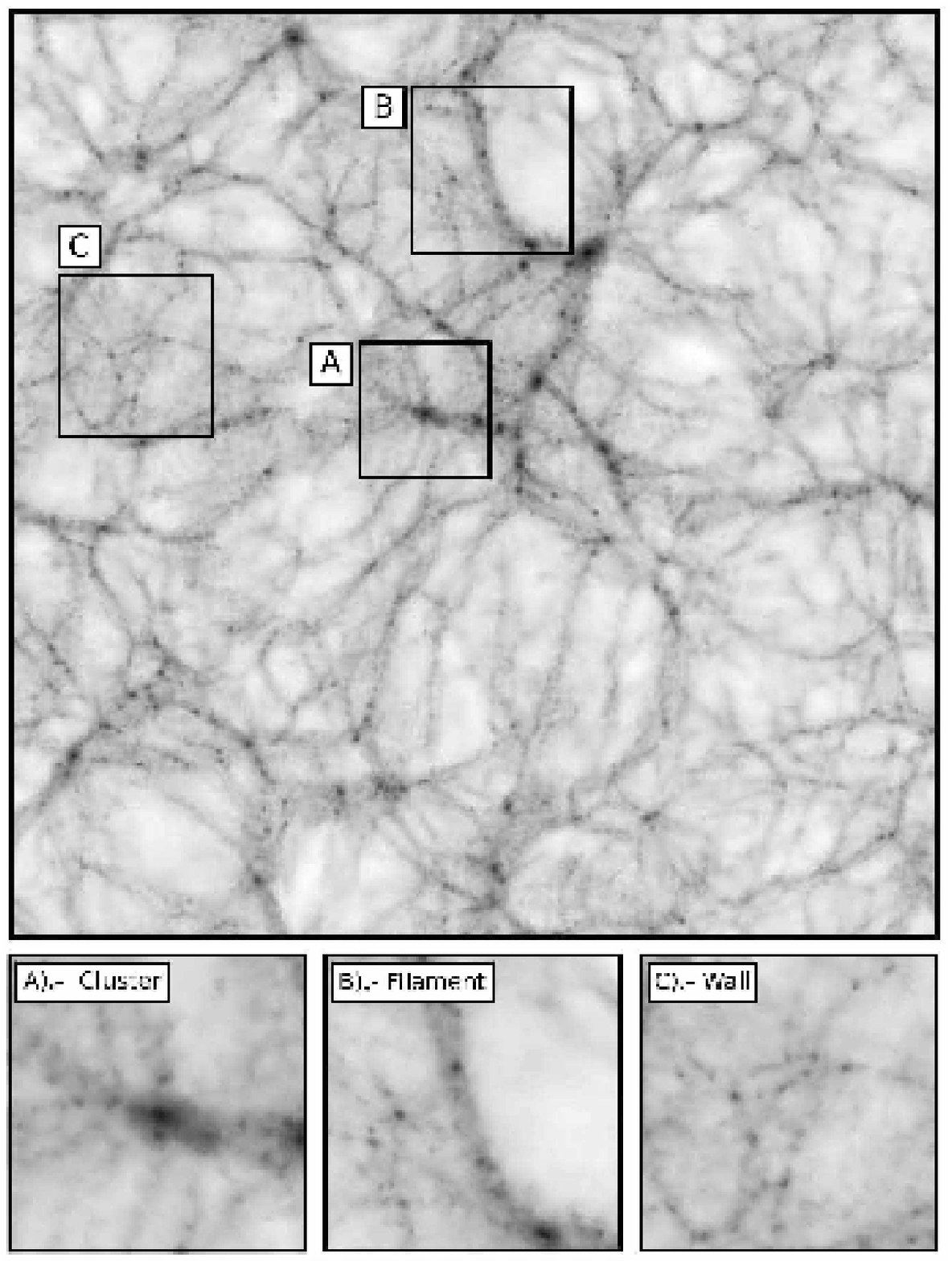}}
  \end{center}
\end{figure*}
\begin{figure*}[t]
  \caption{The Cosmic Web. The image shows the weblike patterns traced by the Dark Matter distribution, 
at the present epoch, in a Universe based on a $\Lambda$CDM scenario. It concerns an N-body simulation in a 
box of $200h^{-1}$ Mpc size. The three boxes indicate examples of the main structure components of the 
Cosmic Web. Amongst others, the image clarifies the mutual spatial relationship between these elements. 
Low-density and low contrast walls are less prominent than the outstanding filamentary channels which define 
the texture of the Cosmic Web. Near the intersection points of filaments and sheets we find high-density 
cluster nodes. The figures demonstrates the significance of the concept ``Cosmic Web''. From 
Arag\'on-Calvo 2007.}
\label{fig:cosmicwebcomponents}
\end{figure*}

\subsection{Dynamical Evolution of the Cosmic Web}
At least three ``universal'' characteristics of the resulting nonlinear cosmic matter 
distribution can be recognized in large N-body simulations of cosmic structure formation such as 
the Millennium simulation \cite{springmillen2005}. One prominent aspect is the {\it hierarchical 
clustering}. The spatial cosmic matter distribution is marked by a large range of scales and 
densities. It is the product of an evolution in which the first objects to condense are small, with 
larger structures forming through the gradual merging of smaller structures. Another prominent 
feature is its {\it anisotropic and weblike spatial geometry}, the consequence of the gravitational 
tendency of overdensities to collapse in an anisotropic fashion. It finds its origin in the intrinsic 
flattening of the overdensities in the primordial density field \cite{bbks}, augmented by the anisotropy of the 
gravitational force field induced by the external matter distribution (i.e. by tidal forces). A third 
important characteristic is the dominant presence of {\it Voids}, large roundish underdense, often near-empty, 
regions. They form in and around density troughs in the primordial density field. They have an essential 
role in the organization of the cosmic matter distribution. Recently, their emergence and evolution has been 
explained within the context of hierarchical gravitational scenarios \cite{shethwey2004}. 

The Cosmic Web theory of Bond et al. \cite{bondweb1996} succeeded in synthesizing all relevant aspects into a coherent dynamical 
and evolutionary framework. Instrumental is the realization that the outline of the cosmic web may already be recognized in 
the primordial density field. The statistics of the primordial tidal field explains why the large scale universe looks 
predominantly filamentary and why in overdense regions sheetlike membranes are only marginal features. Of key importance 
is the realization that the rare high peaks which will eventually emerge as clusters are the dominant agents for generating 
the large scale tidal force field: it is the clusters which weave the cosmic tapestry of filaments 
\cite{bondweb1996,weyedb1996,weygaert2008a}. They cement the structural relations between the 
components of the Cosmic Web and themselves form the junctions at which filaments tie up. This relates the 
strength and prominence of the filamentary bridges to the proximity, mass, shape and mutual orientation of the generating 
cluster peaks: the strongest bridges are those between the richest clusters that stand closely together and point into each 
other's direction. Through the direct relation between the Cosmic Web and the primordial tidal field one may 
understand why the large scale universe looks predominantly filamentary and why in overdense regions 
sheetlike membranes are only marginal features \cite{pogosyan1998}. 

The emerging picture is one of a primordially and hierarchically defined network whose weblike topology is imprinted over a wide 
spectrum of scales. Weblike patterns on ever larger scales get to dominate the density field as cosmic evolution proceeds, 
and as small scale structures merge into larger ones. Within the gradually emptying void regions, however, the topological 
outline of the early weblike patterns remains largely visible.

\begin{figure*}
     \begin{minipage}{\textwidth}
    \begin{center}
      \mbox{\hskip -0.2truecm\includegraphics[width=0.90\textwidth]{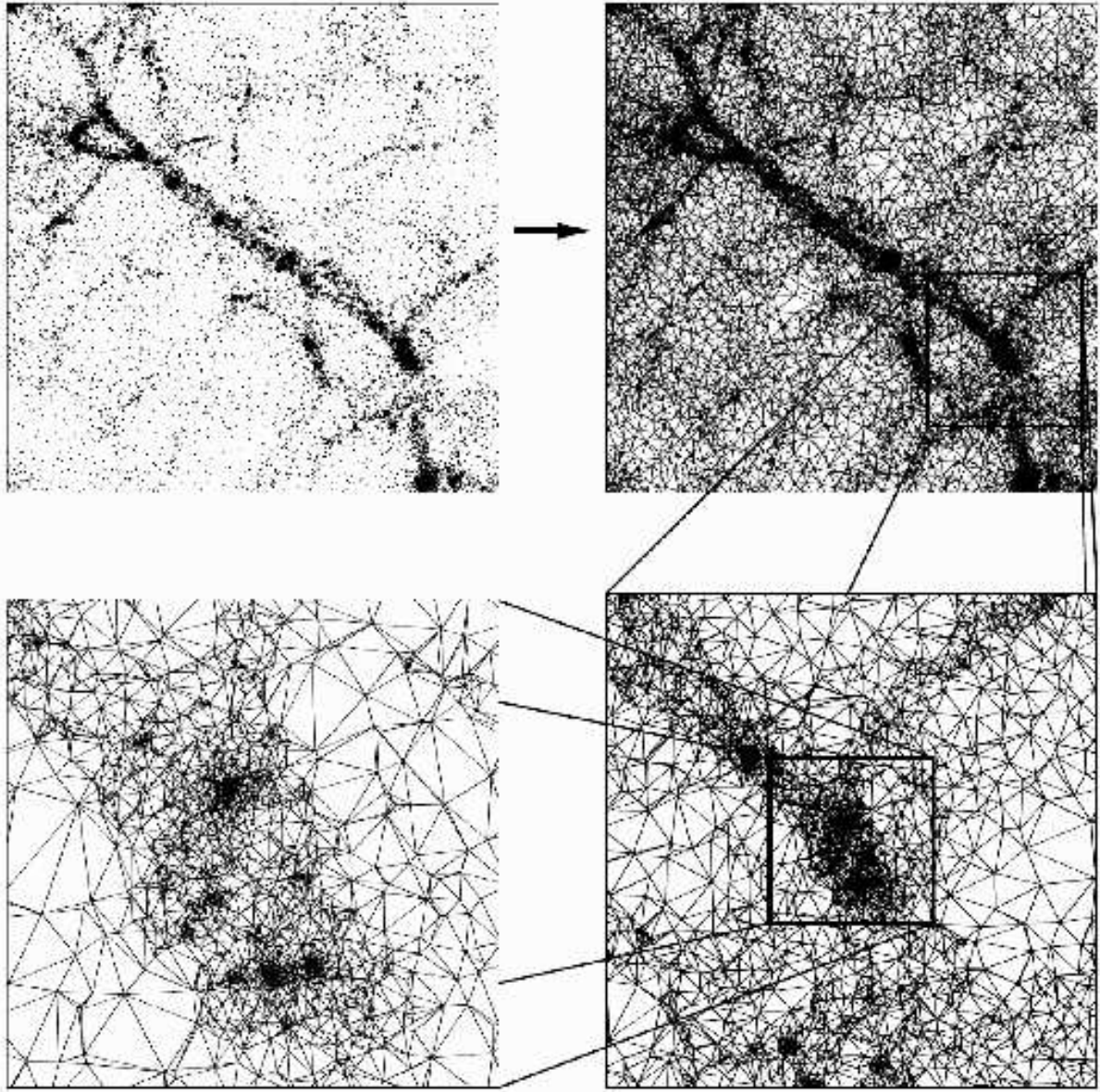}}
    \end{center}
  \end{minipage} 
 \caption{The Delaunay tessellation of a point distribution in and around a filamentary feature. The 
generated tessellations is shown at three successive zoom-ins. The frames form a testimony of the 
strong adaptivity of the Delaunay tessellations to local density and geometry of the spatial point 
distribution. From Schaap 2007.}
  \label{fig:delpntadapt}
\end{figure*} 
\medskip

\subsection{Tessellations and Web Analysis}
Despite a large variety of attempts, as yet no generally accepted descriptive framework has 
emerged for the objective and quantitative analysis of the Cosmic Web. Despite the multitude 
of elaborate qualitative descriptions it has remained a major challenge to characterize its structure, 
geometry and topology. Many attempts to describe, let alone identify, the features and components of the 
Cosmic Web have been of a rather heuristic nature. The overwhelming complexity of both the individual structures 
as well as their connectivity, the lack of structural symmetries, its intrinsic multiscale nature and the wide range 
of densities that one finds in the cosmic matter distribution has prevented the use of simple and straightforward 
toolboxes. 

In the observational reality galaxies are the main tracers of the cosmic web and it is mainly through 
the measurement of the redshift distribution of galaxies that we have been able to map its structure. Likewise, 
simulations of the evolving cosmic matter distribution are almost exclusively based upon N-body particle computer 
calculations, involving a discrete representation of the features we seek to study. Both the galaxy distribution 
as well as the particles in an N-body simulation are examples of {\it spatial point processes} in that they 
are {\it discretely sampled} and have an {\it irregular spatial distribution}.

For furthering our understanding of the Cosmic Web, and to investigate its structure and dynamics, it is of prime importance 
to have access to a set of proper and objective analysis tools. In this contribution we follow the finding that spatial 
tessellations -- in particular Voronoi and Delaunay tessellations -- generated by the discretely sampled cosmic density and 
velocity fields form an ideal basis for an elaborate toolset that succesfully deals with several challenging aspects 
of the analysis of the Cosmic Web:
\begin{itemize}
\item[$\bullet$] Tessellations may be used to interpolate a sample of discrete and irregularly distributed 
values into a {\it volume-weighted} and {\it volume-covering} continuous field.  
\item[$\bullet$] In case the point sample is a representative reflection of an underlying smooth and continuous 
density/intensity field, Voronoi tessellations can be used to estimate and reconstruct the density field. 
\item[$\bullet$] Tessellations are highly sensitive to the morphology of the spatial point distribution. As a 
result they sensitively probe the key aspects - hierarchical matter distribution, anisotropic patterns, 
and voids -- of the Cosmic Web without resorting to user-defined parameters or functions, and without affecting any 
of the other essential characteristics. 
\end{itemize}
\noindent These issues are all addressed by the Delaunay Tessellation Field Estimator (DTFE). The DTFE technique, 
developed by Schaap \& van de Weygaert \cite{schaapwey2000}, forms an elaboration of the velocity interpolation scheme introduced by 
Bernardeau \& van de Weygaert \cite{bernwey1996} towards a multidimensional and fully adaptive scheme for the estimation and 
interpolation of density, velocity and other non-uniformly and discretely sampled quantities to yield a corresponding 
volume-covering continuous field. 

The application of DTFE to the observed spatial distribution of galaxies, or that of particles in 
a computer N-body simulation, produces a continuous density and velocity field which retains the 
crucial aspects of its hierarchical nature and anisotropic morphology. This may be directly appreciated 
from the three consecutive zoom-ins of the DTFE rendered cosmic web density field in 
fig.~\ref{fig:hierarchydtfe}, and its comparison to the more conventional gridbased TSC 
interpolation scheme.  Because of these characteristics, the DTFE density field forms the basis for a diverse 
set of tools for the analysis of various aspects of the cosmic matter distribution. In this contribution we 
will report on our work on the Multiscale Morphology Filter for detecting filaments and sheets, the Watershed 
Void Finder for detecting voids, the Cosmic Spine formalism for the analysis of the filamentary network of the 
Cosmic Web and, finally, the related Alphashape analysis of the topological structure of the cosmic matter distribution. 

\section{DTFE: fundamentals}
\label{sec:dtfefund}
\noindent DTFE obtains optimal {\it local} estimates of the spatial density (see \cite{okabe2000}, sect.~8.5), while the tetrahedra of its 
{\it dual} Delaunay tessellation are used as multidimensional intervals for {\it linear} interpolation of the field values sampled 
or estimated at the location of the sample points (see \cite{okabe2000}, ch. 6). With respect to the aspect of interpolation, 
the Delaunay Tessellation Field Estimator is the linear version of nn-neighbour interpolation techniques (see sect.~\ref{sec:nn}). 
Its core, however, is the additional aspect of being able to translate the generating spatial point distribution into a continuous 
density field. 

The DTFE method has been first defined in the context of a description and analysis of cosmic flow fields which are sampled 
by a set of discretely and sparsely sampled galaxy peculiar velocities. Bernardeau \& van de Weygaert \cite{bernwey1996} demonstrated 
the method's superior performance with respect to conventional interpolation procedures. They also proved that the obtained field 
estimates involve those of the proper {\it volume-weighted} quantities, instead of the usually implicit {\it mass-weighted} quantities (see 
sect.~\ref{sec:massvolweight}). This corrected a few fundamental biases in estimates of higher order 
velocity field moments.

The DTFE technique allows us to follow the same geometrical and structural adaptive properties of the higher 
order nn-neighbour methods while allowing the analysis of truely large data sets and include the reconstruction 
of the most important aspect, that of the cosmological density field. For three-dimensional samples with large 
number of points, akin to those found in large cosmological computer simulations, the more complex geometric 
operations involved in the pure nn-neighbour interpolation still represent a computationally challenging task 
(see sect.~\ref{sec:nn}). To deal with the large point samples consisting of hundreds of thousands to several millions of points we chose 
to follow a related nn-neigbhour based technique that restricts itself to pure linear interpolation. 

\subsection{Voronoi and Delaunay Adaptivity}
\noindent The primary ingredient of the DTFE procedure, and the related Natural Neighbour Interpolation methods, is the Delaunay 
tessellation of the particle distribution ${\mathcal P}$. The Delaunay and Voronoi cells adjust themselves to the 
spatial characteristics of the point distribution. This concerns its spatial resolution and its geometry. 

DTFE, and its higher-order equivalent {\it Natural Neigbhour Interpolation} \cite{sibson1980,sibson1981,watson1992,braunsambridge1995,
sukumarphd1998}, exploit three properties of Voronoi and Delaunay tessellations (see eg. \cite{schaapphd2007,weyschaap2007}): 
\begin{itemize}
\item[$\bullet$] The tessellations are very sensitive to the local point density. DTFE uses this to define a local 
estimate of the density on the basis of the inverse of the volume of the tessellation cells. 
\item[$\bullet$] The sensitivity of Voronoi and Delaunay tessellations to the local geometry of the point distribution. This allows 
DTFE to accurately trace anisotropic features, such as seen in the Cosmic Web.
\item[$\bullet$] The adaptive and minimum triangulation properties of Delaunay tessellations allow them to be used 
as adaptive spatial interpolation intervals for irregular point distributions. This has also been recognized in 
a 
\end{itemize}

In sparsely sampled regions the Voronoi and Delaunay cells are large and the distance between {\it natural neighbours}, points 
that share a Voronoi simplex, is large. Not only the size, but also the shape of the Delaunay tetrahedra is fully determined by 
the spatial point distribution: if is anisotropic this will be reflected in the distribution. In fig.~\ref{fig:delpntadapt} we 
see that  the Delaunay tessellation traces the density and geometry of the local point distribution at three consecutive 
spatial scales  to a remarkable degree. The density and geometry of the local point distribution will therefore determine
the resolution of the spatial interpolation and reconstruction procedures based on the Voronoi and Delaunay tessellation. 

\subsection{Natural Neighbour Interpolation}
\label{sec:nn}
\noindent {\it Natural Neighbour Interpolation} formalism is a generic higher-order multidimensional interpolation, smoothing and modelling 
procedure utilizing the concept of natural neighbours to obtain locally optimized measures of system characteristics. Its theoretical 
basis was developed and introduced by Sibson \cite{sibson1981}, while extensive treatments and elaborations of nn-interpolation may be 
found in \cite{watson1992,sukumarphd1998}. Natural neighbour interpolation produces a conservative, artifice-free, result 
by finding area-weighted weighted averages, at each interpolation point, of the functional values associated with that subset of 
data which are natural neighbors of each interpolation point. According to the nn-interpolation scheme the interpolated value 
${\widehat f}({\bf x})$ at a position ${\bf x}$ is given by 
\begin{equation}
{\widehat f}({\bf x})\,=\,\sum_i\,\phi_{nn,i}({\bf x})\,f_i\,,
\label{eq:nnint}
\end{equation}
in which the summation is over the natural neighbours of the point ${\bf x}$, i.e. the 
sample points $j$ with whom the order-2 Voronoi cells ${\mathcal V}_2({\bf x},{\bf x}_j)$ are 
not empty. Sibson interpolation is based upon 
the interpolation kernel $\phi({\bf x},{\bf x}_j)$ to be equal to the normalized order-2 Voronoi cell, 
\begin{equation}
\phi_{nn,i}({\bf x})\,=\,{\displaystyle {{\mathcal A}_{2}({\bf x},{\bf x}_i)} \over 
{\displaystyle {\mathcal A}({\bf x})}}\,,
\label{eq:nnintint}
\end{equation}
in which ${\mathcal A}({\bf x})=\sum_j {\mathcal A}({\bf x},{\bf x}_j)$ is the volume of the 
potential Voronoi cell of point ${\bf x}$ if it had been added to the point sample ${\mathcal P}$ and  
the volume ${\mathcal A}_{2}({\bf x},{\bf x}_i)$ concerns the order-2 Voronoi cell 
${\mathcal V}_2({\bf x},{\bf x}_i)$, the region of space for which the points ${\bf x}$ and ${\bf x_i}$ 
are the closest points. The interpolation kernels $\phi$ are always positive and sum to one. The resulting 
function is continuous everywhere within the convex hull of the data, and has a continuous slope everywhere 
except at the data themselves. 

\subsection{DTFE Interpolation}
\noindent The linear interpolation scheme of DTFE exploits the same spatially adaptive characteristics of the Delaunay tessellation 
generated by the point sample ${\mathcal P}$ as that of regular natural neighbour schemes. For DTFE the interpolation kernel 
$\phi_{dt,i}({\bf x})$ is that of regular linear interpolation within the Delaunay tetrahedron in which ${\bf x}$ is located 
(see eq.~\ref{eq:fieldval}),
\begin{equation}
{\widehat f}_{dt}({\bf x})\,=\sum_i\,\phi_{dt,i}({\bf x})\,f_i,
\label{eq:dtfeint}
\end{equation}
in which the sum is over the four sample points defining the Delaunay tetrahedron. Note that for both the nn-interpolation as 
well as for the linear DTFE interpolation, the interpolation kernels $\phi_i$ are unity at sample point location ${\bf x}_i$ and 
equal to zero at the location of the other sample points $j$,
\begin{equation}
\phi_i({\bf x}_j)\,=\,
\begin{cases}
1\hskip 2.truecm {\rm if}\,i=j\,,\\
0\hskip 2.truecm {\rm if}\,i\ne j\,,
\end{cases}\,
\end{equation}
where ${\bf x}_j$ is the location of sample point $j$.

In practice, it is convenient to replace eqn.~\ref{eq:dtfeint} with its equivalent expression 
in terms of the (linear) gradient ${\widehat {\nabla f}}\bigr|_m$ inside the Delaunay simplex $m$,
\begin{equation}
{\widehat f}({\bf x})\,=\,{\widehat f}({\bf x}_{i})\,+\,{\widehat {\nabla f}} \bigl|_m \,\cdot\,({\bf x}-{\bf x}_{i}) \,.
\label{eq:fieldval}
\end{equation}
\noindent The value of ${\widehat {\nabla f}}\bigr|_m$ can be easily and uniquely determined from the $(1+D)$ field values $f_j$ at 
the sample points constituting the vertices of a Delaunay simplex. Given the location ${\bf r}=(x,y,z)$ of the four 
points forming the Delaunay tetrahedra's vertices, ${\bf r}_0$, ${\bf r}_1$, ${\bf r}_2$ and ${\bf r}_3$, 
and the value of the sampled field at each of these locations, $f_0$, $f_1$, $f_2$ and $f_3$ and defining the 
quantities 
\begin{eqnarray}
\Delta x_n &\,=\,& x_n-x_0\,;\,\nonumber\\
\Delta y_n &\,=\,& y_n -y_0\,;\quad \hbox{for}\ n=1,2,3\\
\Delta z_n &\,=\,& z_n -z_0\,,\nonumber 
\end{eqnarray}
\noindent as well as $\Delta f_n\,\equiv\,f_n-f_0\,(n=1,2,3)$ the gradient $\nabla f$ follows from the inversion
\begin{eqnarray}
\nabla f&\,=\,&
\begin{pmatrix}
{\displaystyle \partial f \over \displaystyle \partial x}\\
\ \\
{\displaystyle \partial f \over \displaystyle \partial y}\\
\ \\
{\displaystyle \partial f \over \displaystyle \partial z}
\end{pmatrix}
\,=\,{\bf A}^{-1}\,
\begin{pmatrix}
\Delta f_{1} \\ \ \\ \Delta f_{2} \\ \ \\ \Delta f_{3} \\ 
\end{pmatrix}\,;\qquad\nonumber\\
\ \\
{\bf A}&\,=\,&
\begin{pmatrix}
\Delta x_1&\Delta y_1&\Delta z_1\\
\ \\
\Delta x_2&\Delta y_2&\Delta z_2\\
\ \\
\Delta x_3&\Delta y_3&\Delta z_3
\end{pmatrix}\nonumber
\label{eq:fieldgrad}
\end{eqnarray}
\noindent Once the value of $\nabla_f$ has been determined for each Delaunay tetrahedron in the tessellation, it is 
straightforward to determine the DTFE field value ${\widehat f}({\bf x})$ for any location ${\bf x}$ by means 
of straightforward linear interpolation within the Delaunay tetrahedron in which ${\bf x}$ is located 
(eqn.~\ref{eq:fieldval}).

The one remaining complication is to locate the Delaunay tetrahedron ${\mathcal D}_m$ in which a particular 
point ${\bf x}$ is located. This is not as trivial as one might naively think. It not necessarily concerns 
a tetrahedron of which the nearest nucleus is a vertex. Fortunately, a very efficient method, the {\it 
walking triangle algorithm}~\cite{lawson1977,sloan1987} has been developed. Details of the method may be 
found in \cite{sbm1995,schaapphd2007}. 

\subsection{DTFE Density Estimates}
\noindent The DTFE procedure extends the concept of interpolation of field values sampled at the point sample ${\mathcal P}$ 
to the estimate of the density ${\widehat \rho}({\bf x})$ from the spatial point distribution itself. This is only 
feasible if the spatial distribution of the discrete point sample forms a fair and unbiased reflection of the 
underlying density field. 

It is commonly known that an optimal estimate for the spatial density at the location of a point ${\bf x}_i$ in a 
discrete point sample ${\mathcal P}$ is given by the inverse of the volume of the corresponding Voronoi cell (see 
\cite{okabe2000}, for references). Tessellation-based methods for estimating the density have been introduced by Brown \cite{brown1965} 
and Ord \cite{ord1978}. In astronomy, Ebeling \& Wiedenmann \cite{ebeling1993} were the first to use tessellation-based density 
estimators for the specific purpose of devising source detection algorithms. This work has recently been applied to cluster detection 
algorithms by \cite{ramella2001,kim2002,marinoni2002,lopes2004,neyrinck2005}. Along the same lines, Ascasibar \& Binney 
\cite{ascalbin2005} suggested 
that the use of a multidimensional binary tree might offer a computationally more efficient alternative. However, these studies 
have been restricted to raw estimates of the local sampling density at the position of the sampling points and have not yet 
included the more elaborate interpolation machinery of the DTFE and Natural Neighbour Interpolation methods. 

\begin{figure*}
  \centering
    \mbox{\hskip -0.25truecm\includegraphics[width=0.98\textwidth]{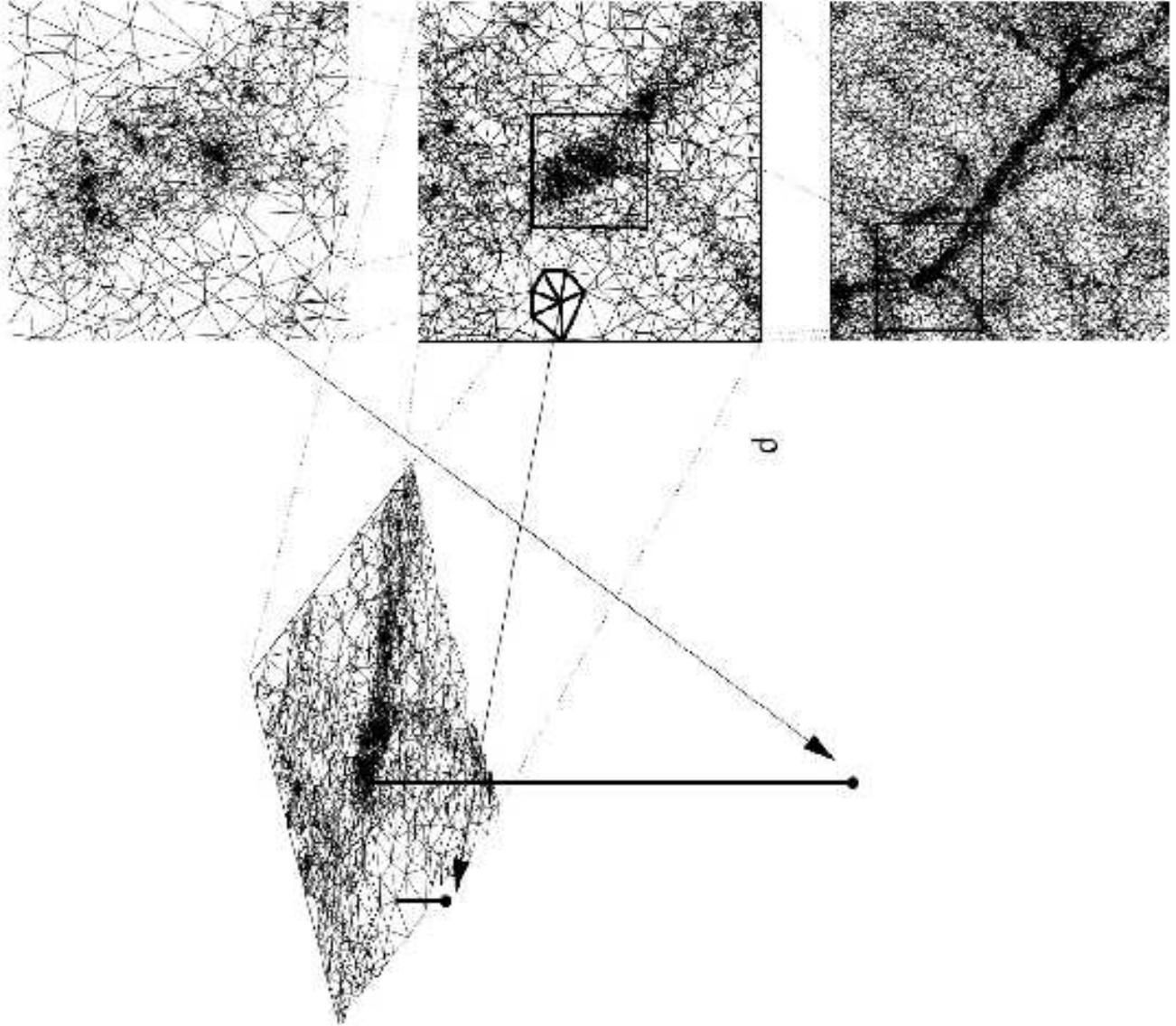}}
    \caption{Relation between density and volume contiguous Voronoi cells. The figure (top row) zooms in 
on the Delaunay tessellation defined by the particle distribution in and around a filament. The density 
values at two sample points within the filamentary structure are indicated (bottom panel). From Schaap 2007.}
    \vskip -0.5truecm
\label{fig:dtfestep2} 
\end{figure*} 

\begin{figure*}
\begin{center}
\vskip -0.5truecm
\mbox{\hskip -0.6truecm\includegraphics[width=0.98\textwidth]{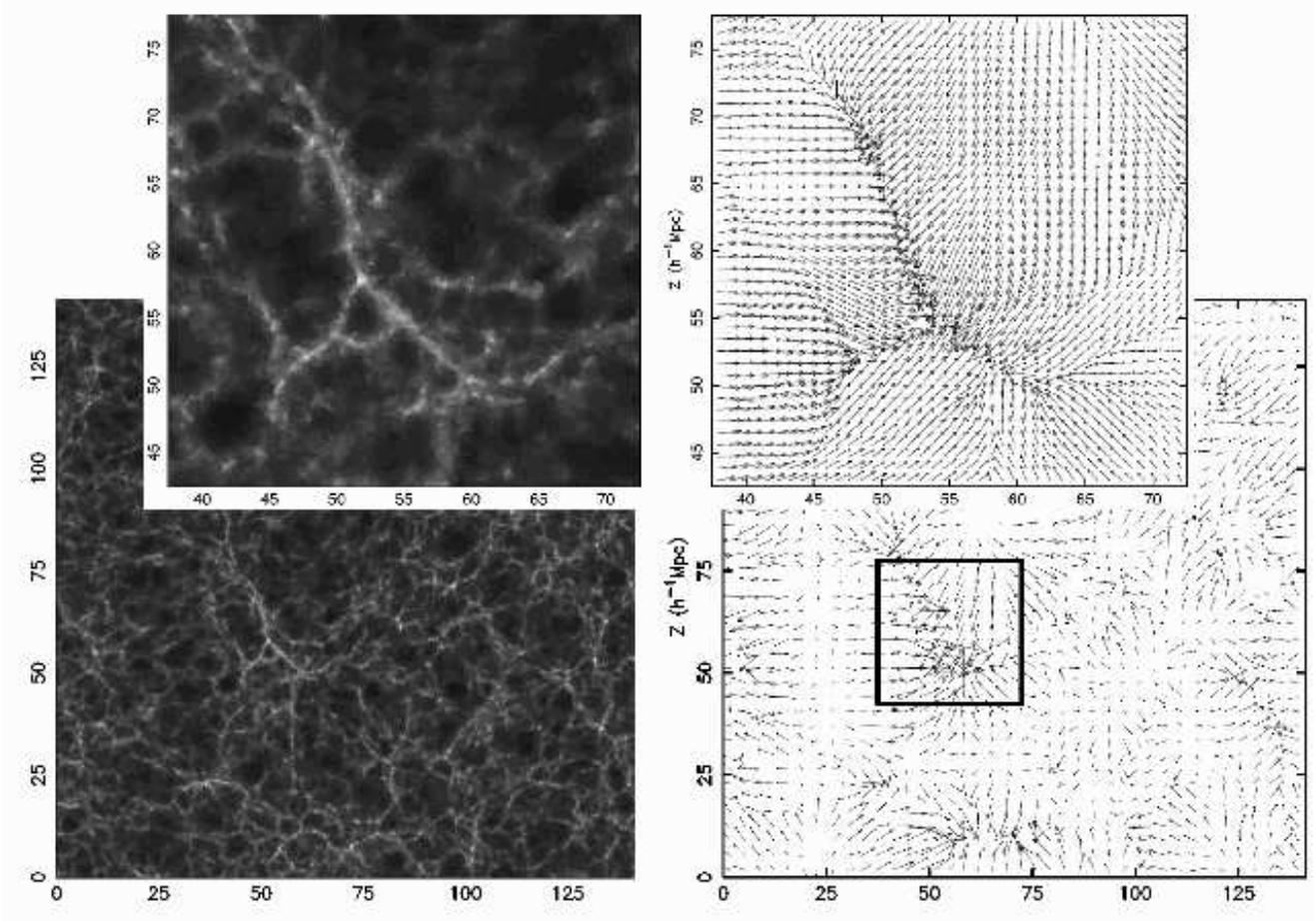}}
\vskip 0.0truecm
\caption{The density and velocity field of the LCDM GIF N-body simulation, by means of DTFE 
zooming in on the flow field around a filamentary structure. The shear 
flow along the filaments is meticulously resolved.}
\label{fig:giffil}
\end{center}
\end{figure*}

The density field reconstruction of the DTFE procedure consists of two steps, the zeroth-order estimate ${\widehat \rho}_0$ of the 
density values at the location of the points in ${\mathcal P}$ and the subsequent linear interpolation of those 
zeroth-order density estimates over the corresponding Delaunay grid throughout the sample volume. This 
yields the DTFE density field estimate ${\widehat \rho}({\bf x})$. 

An {\it essential} requirement for cosmological purposes of our interpolation scheme is that 
the estimated DTFE density field ${\widehat \rho}({\bf x})$ should guarantee {\it mass conservation}: 
the total mass corresponding to the density field should be equal to the mass represented by the 
sample points. Indeed, this is an absolutely crucial condition for many applications of a 
physical nature. Since the mass $M$ is given by the integral of the density field $\rho({\bf x})$ over 
space, this translates into the integral requirement 
\begin{eqnarray}
{\widehat M}&\,=\,&\int {\widehat \rho}({\bf x})\,d{\bf x}\nonumber\\
&\,=\,&\sum_{i=1}^{N} m_i\,=\,M\,=\,{\rm cst.},
\label{eq:integratemass}
\end{eqnarray}
with $m_i=m$ is the mass per sample point. It is straightforward to infer that if the zeroth-order 
estimate of the density values would be the inverse of the regular Voronoi volume the condition 
of mass conservation would not be met. Instead, the DTFE procedure employs a slightly modified 
yet related zeroth-order density estimate, the normalized inverse of the volume $V({\mathcal W}_i)$ 
of the {\it contiguous Voronoi cell} ${\mathcal W}_i$ of each point $i$. For $D$-dimensional 
space this is 
\begin{equation}
{\widehat \rho}({\bf x}_i)\,=\,(1+D)\,\frac{m_i}{V({\mathcal W}_i)} \,.
\label{eq:densdtfe}
\end{equation}
The {\it contiguous Voronoi cell} of a point $i$ is the union of all Delaunay tetrahedra 
of which point $i$ forms one of the four vertices. It is straightforward to prove that with 
the contiguous Voronoi cell density estimator mass conservation is guaranteed (see \cite{weyschaap2007}, sect. 10). 

\subsection{Volume-weighted and Mass-weighted fields}
\label{sec:massvolweight}
For relating measured quantities to theoretical predictions, one has to take 
account of the implicit filtering process involved in translating measurements 
(including that in computer simulations) to interpretable measures. When dealing 
with a probe of an underlying density field $\rho({\bf x})$ - e.g. the galaxy distribution in 
a cosmological context - there are two possibilities. Theoretically the cleanest measure is 
that of the volume-weighted quantity $f_{vol}$,
\begin{equation}
{\widehat f}_{vol}({\bf x})\,\equiv\,{\displaystyle \int\,d{\bf y}\, f({\bf y})\,W({\bf x}-{\bf y})
\over \displaystyle \int\,d{\bf y}\,\,W({\bf x}-{\bf y})}\,.
\label{eq:fvol}
\end{equation}
In practice, however, most techniques implicitly (and often unknowingly) yield the 
{\it mass-weighted} field averages. 
\begin{equation}
{\widehat f}_{mass}({\bf x})\,\equiv\,{\displaystyle \int\,d{\bf y}\, f({\bf y})\,\rho({\bf y})\,W({\bf x}-{\bf y})
\over \displaystyle \int\,d{\bf y}\,\rho({\bf y})\,W({\bf x}-{\bf y})}
\label{eq:fmass}
\end{equation}
\noindent where $W({\bf x},{\bf y})$ is the adopted filter function defining the weight of a mass element in a way 
that is dependent on its position $y$ with respect to the position ${\bf x}$. It turns out that the 
use of the Voronoi and Delaunay tessellation guarantees the {\it volume-weighted} nature of the 
DTFE fields \cite{bernwey1996}. 

An additional crucial ingredient of any reconstruction procedure are its error 
characteristics. Within the limited context of the present contribution it is not feasible 
to include a detailed discussion of the noise and error characteristics of 
DTFE reconstructed density and velocity fields. For this we refer the reader 
elsewhere \cite{schaapphd2007,weyschaap2007}.

\section{the DTFE procedure}
\begin{figure*}
  \centering
    \mbox{\hskip -0.25truecm\includegraphics[height=0.99\textwidth,angle=90.0]{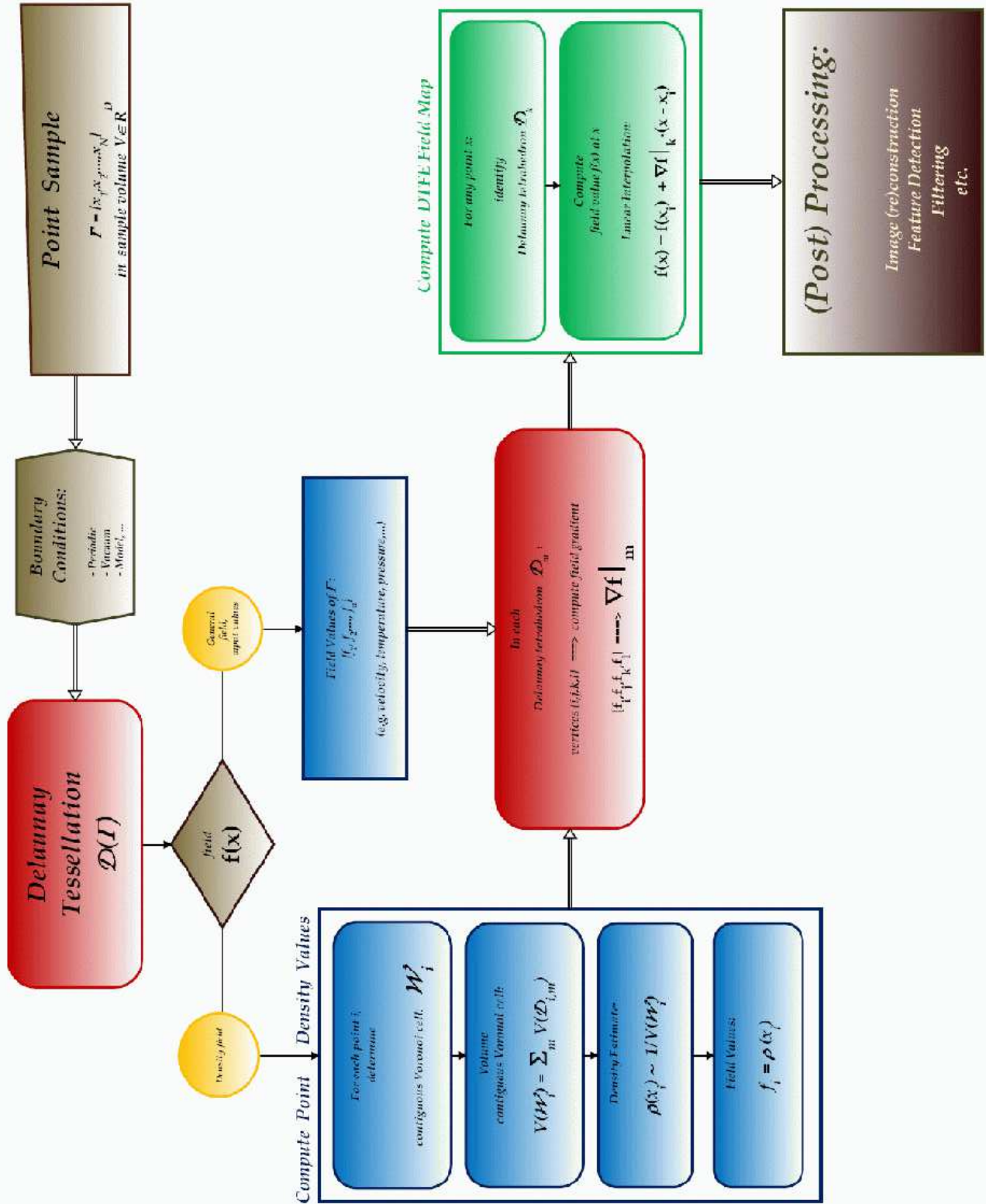}}
    \caption{Schematic diagram of the DTFE procedure.}
    \vskip -0.5truecm
\label{fig:dtfediag} 
\end{figure*}

\noindent The complete DTFE reconstruction procedure is shown in the 
the schematic diagram of fig.~\ref{fig:dtfediag}. It involves the following 
steps:
\medskip
\begin{enumerate}
\item[$\bullet$] {\bf Point sample}\\
Defining the spatial distribution of the point sample.
\item[$\bullet$] {\bf Delaunay Tessellation}\\
Construction of the Delaunay tessellation from the point sample. 
\item[$\bullet$] {\bf Field values point sample}\\
Dependent on whether it concerns the densities at the sample points or a 
measured field value there are two options:
\begin{enumerate}
\item[+] {\it General (non-density) field}:\ \\ (Sampled) value of field at sample point.
\item[+] {\it Density field}
\end{enumerate}
\item[$\bullet$] {\bf Field Gradient}\\
 Calculation of the field gradient estimate $\widehat{\nabla f}|_m$ 
 in each $D$-dimensional Delaunay simplex $m$ ($D=3$: tetrahedron; $D=2$: triangle) 
by solving the set of linear equations for the field values at the positions 
of the $(D+1)$ tetrahedron vertices,
\begin{equation}
\widehat{\nabla f}|_m \ \ \Longleftarrow\ \ 
\begin{cases}
f_0 \ \ \ \ f_1 \ \ \ \ f_2 \ \ \ \ f_3  \\
\ \\
{\bf r}_0 \ \ \ \ {\bf r}_1 \ \ \ \ {\bf r}_2 \ \ \ \ {\bf r}_3 
\end{cases}\,
\label{eq:dtfegrad}
\end{equation}

\bigskip
\item[$\bullet$] {\bf Interpolation}.\\
The final basic step of the DTFE procedure is the field interpolation. The processing 
and postprocessing steps involve numerous interpolation calculations, for each of the 
involved locations ${\bf x}$. Given a location ${\bf x}$, the Delaunay tetrahedron $m$ in which it is embedded is 
determined. On the basis of the field gradient $\widehat{\nabla f}|_m$ the field value 
is computed by (linear) interpolation, 
\begin{equation}
{\widehat f}({\bf x})\,=\,{\widehat f}({\bf x}_{i})\,+\,{\widehat {\nabla f}} \bigl|_m \,\cdot\,({\bf x}-{\bf x}_{i}) \,.
\end{equation}
In principle, higher-order interpolation procedures are also possible.

\bigskip
\item[$\bullet$] {\bf Processing}.\\
Though basically of the same character for practical purposes we make a distinction between 
straightforward processing steps concerning the production of images and simple smoothing 
filtering operations on the one hand, and more complex postprocessing on the other hand. 
The latter are treated in the next item. Basic to the processing steps is the determination 
of field values following the interpolation procedure(s) outlined above.\\ 
Straightforward ``first line'' field operations are {\it ``Image reconstruction''} and, 
subsequently, {\it ``Smoothing/Filtering''}.
\begin{enumerate}
\item[+] {\it Image reconstruction}.\\ For a set of {\it image points}, usually grid points, 
determine the {\it image value}: formally the average field value within the corresponding gridcell.
\item[+] {\it Smoothing} and {\it Filtering}:
\end{enumerate}

\item[$\bullet$] {\bf Post-processing}.\\
The real potential of DTFE fields may be found in sophisticated applications, 
tuned towards uncovering characteristics of the reconstructed fields. 
An important aspect of this involves the analysis of structures in the 
density field. Some notable examples are:
  \begin{enumerate}
    \item[+] Advanced filtering operations. Potentially interesting 
applications are those based on the use of wavelets \cite{martinez2005}.
    \item[+] Cluster, Filament and Wall detection by means of the 
{\it Multiscale Morphology Filter} \cite{miguelphd2007,aragonmmf2007}.
    \item[+] Void identification on the basis of the {\it cosmic watershed} 
algorithm~\cite{platen2007,neyrinck2008}.
    \item[+] Tracing the filamentary spine of the Cosmic Web, 
on the basis of a Morse-theory related analysis, the {\it Cosmic Spine} 
\cite{aragonspine2009a}.
    \item[+] Halo detection in N-body simulations \cite{neyrinck2005}.
    \item[+] The computation of 2-D surface densities for the study of 
gravitational lensing \cite{bradac2004}.
  \end{enumerate} 
\end{enumerate}

\begin{figure*}
  \center
     \mbox{\hskip -0.5truecm\includegraphics[width=0.90\textwidth]{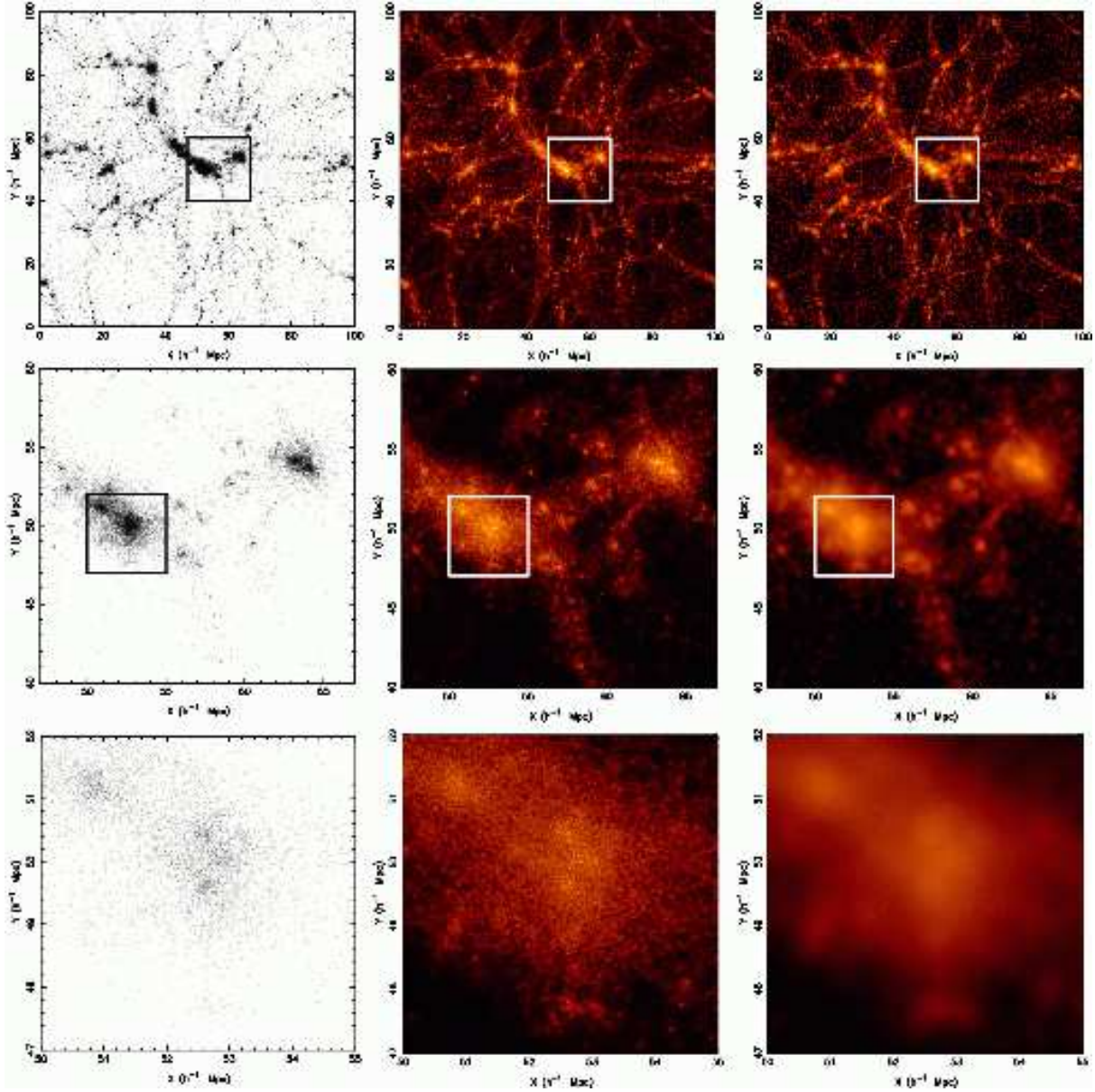}}
  \vskip -0.2truecm
  \caption{DTFE cosmic density field illustrating the large dynamic range 
  which is present in the large scale matter distribution, and the ability of DTFE to 
  reproduce the key characteristics of the Cosmic Web. Top: 
  cosmic web dark matter density field in a 10$h^{-1}$Mpc wide slice 
  through a $\Lambda$CDM computer simulation of cosmic structure formation. 
  In the subsequent frames zoom-ins focus in on a high-density supercluster 
  region around a massive cluster (zoom-in boxes indicated). }
\label{fig:hierarchydtfe}
\end{figure*}

\begin{figure*}
  \centering
    \mbox{\hskip -0.7truecm\includegraphics[width=0.90\textwidth]{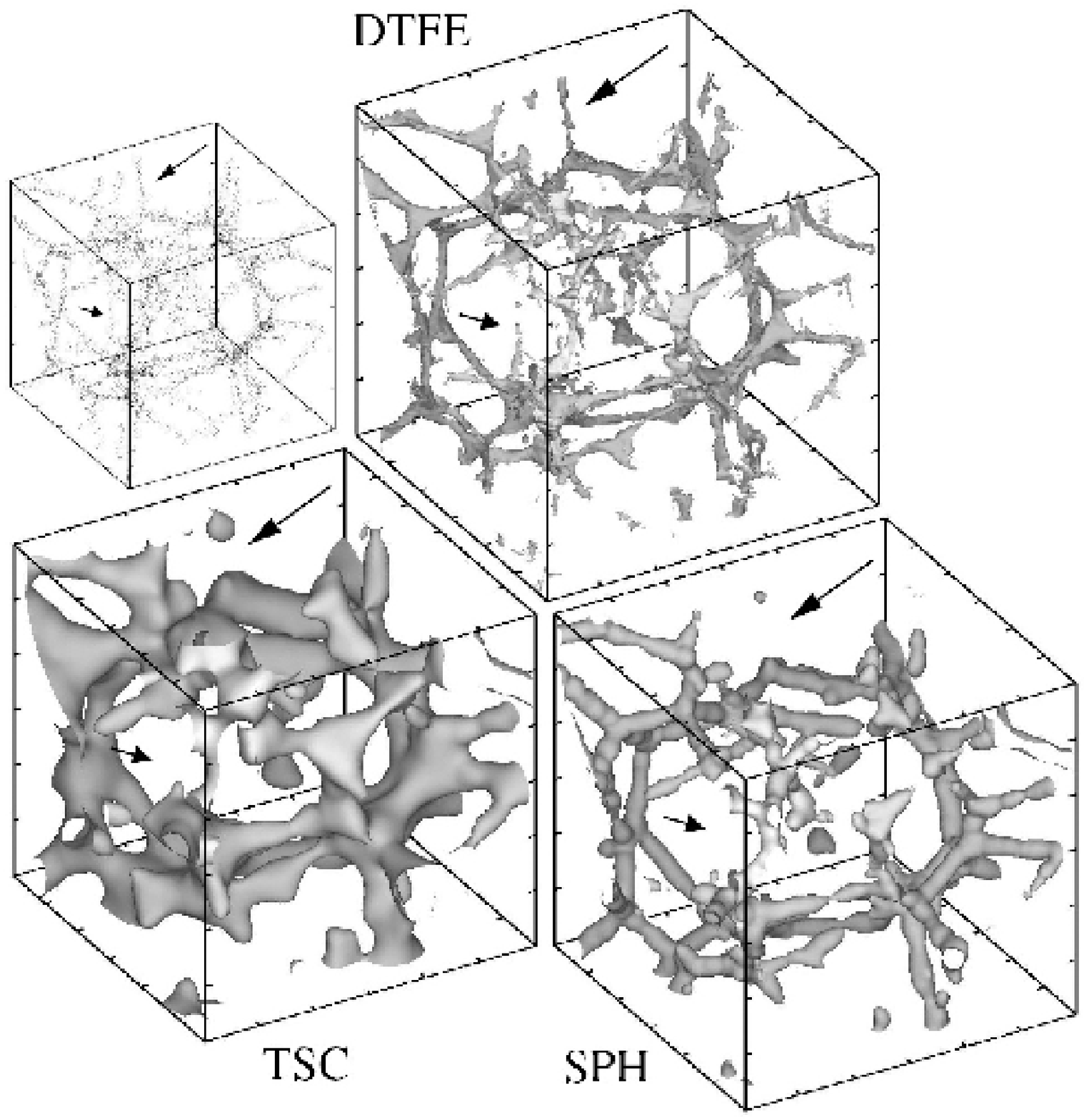}}
    \vskip -0.0truecm
    \caption{Three-dimensional visualization of a filamentary particle 
distribution. The DTFE reconstruction delineates the full weblike network. 
The density contours have been chosen such that $65\%$ of the mass is enclosed. The 
reconstructions on the basis of the TSC and SPH spline interpolations. The arrows indicate two
     structures which are visible in both the galaxy distribution and
     the DTFE reconstruction, but not in the TSC and SPH spline-based reconstructions.}.
     \vskip -0.25truecm
\label{fig:vorfil3d} 
\end{figure*} 
\bigskip
\noindent In addition, DTFE enables the simultaneous and combined analysis of density 
fields and other relevant physical fields. As it allows the simultaneous 
determination of {\it density} and {\it velocity} fields, it can serve as 
the basis for studies of the dynamics of structure formation in the cosmos. 
Its ability to detect substructure as well as reproduce the morphology 
of cosmic features and objects implies DTFE to be suited for assessing their
dynamics without having to invoke artificial filters.\\
\medskip

\section{DTFE and the Cosmic Web}
\noindent Figure~\ref{fig:giffil} provides a nice impression of the versatility of the DTFE formalism.
DTFE density and velocity fields may be depicted at any arbitrary resolution without involving 
any extra calculation: zoom-ins represent themselves a real magnification of the reconstructed fields. 
This is in stark contrast to conventional reconstructions in which the resolution is arbitrarily set 
by the users and whose properties are dependent on the adopted resolution.

It shows the density and velocity field, at two resolutions, in and around filaments and 
other components of the cosmic web in a GIF $\Lambda$CDM simulation. DTFE not only 
it allows a study of the patterns in the nonlinear matter distribution but also a study of the 
related velocity flows. Because DTFE manages to follow both the density distribution and the 
corresponding velocity distribution into nonlinear features it opens up the window towards 
a study of the dynamics of the formation of the cosmic web and its corresponding elements 
\cite{romwey2007}. Note that such an analysis of the dynamics is limited to regions and scales 
without multistream flows. 

\medskip
\subsection{DTFE characteristics}\ \\
\noindent Within the cosmological context a major -- and crucial -- characteristic of a processed DTFE 
density and velocity field is that it is capable of delineating the three fundamental characteristics of 
the spatial structure of the Megaparsec cosmic matter distribution:
\begin{itemize}
\item[$\bullet$] It outlines the full hierarchy of substructures present in the sampling point distribution, 
relating to the standard view of structure in the Universe having arisen through the gradual hierarchical 
buildup of matter concentrations (fig.~\ref{fig:hierarchydtfe}).  
\item[$\bullet$] DTFE also reproduces any anisotropic patterns in the density distribution without diluting 
their intrinsic geometrical properties. This is particularly important when analyzing the prominent filamentary 
and planar features marking the Cosmic Web (fig.~\ref{fig:vorfil3d}).
\item[$\bullet$] A third important aspect of DTFE is that it outlines the presence and shape of voidlike regions. 
Because of the interpolation definition of the DTFE field reconstruction voids are rendered as regions of slowly 
varying and moderately low density values.
\end{itemize}

\noindent The success of DTFE in following these key characteristics has been substantiated by quantitative tests. 
In \cite{schaapwey2009b} we demonstrate DTFE's ability to infer the fractal dimensions of a range of 
fractal point distributions: DTFE follows the density distribution throughout the available spatial and 
mass range. Additional tests in \cite{schaapwey2009c} have confirmed the ability of DTFE to reconstruct 
density fields with the correct shape morphology (inertia tensor), whether it related to distributions 
dominated by compact high density cluster peaks, elongated filaments or sheets. Its comparative performance 
with respect to other interpolation schemes may be assessed from fig.~\ref{fig:vorfil3d}. Clearly, 
DTFE remains closer to the actual structure represented in the particle distribution. 

In a thorough analysis of the interpolation and morphological performance of DTFE, Platen et al. \cite{platen2009} 
has demonstrated that DTFE produces better and more reliable results than {\it Natural Neighbour interpolation}  
and various {\it Kriging} schemes. Even while the latter do take into account long-range spatial correlations, 
the self-adaptive local character of DTFE appears not only to be better suited for uniformly sampled 
spatial datasets, but also for dealing with datasets with a radially varying selection function or 
datasets beset by systematic redshift distortions due to the peculiar velocities of galaxies\footnote{Redshift 
distortions: in addition to their cosmic expansion velocity, galaxies have velocities with respect to the 
expanding Universe. The Doppler redshift corresponding to these velocities add to the cosmic redshift to 
yield a redshift $z$ that is not precisely proportional to the distance $r$ of an object.}.

\subsection{Analysis of the Cosmic Web}
\noindent Within its cosmological context, DTFE will meet its real potential in more sophisticated applications tuned 
towards uncovering morphological characteristics of the reconstructed spatial patterns. The true potential of DTFE and 
related adaptive random tessellation based techniques comes to the fore when we take DTFE reconstructions as the basis 
for a variety of ``post-processing'' tools. A variety of recent techniques have recognized the high dynamic range and 
adaptivity of tessellations to the spatial and morphological resolution of the systems they seek to analyze. 

We have been working on three different, yet mutually related, formalisms which use DTFE density fields as the basis for 
the identification of various key aspects of the Cosmic Web. The Multiscale Morphology Filter (MMF), introduced and 
defined by Arag\'on-Calvo et al. \cite{aragonmmf2007}, attempts to detect weblike anisotropic features over a range of spatial scales and 
implicitly incorporates the hierarchical nature of the Megaparsec matter distribution. While incorporating the use 
of a more or less artificial scale filter, the {\it Watershed Void Finder} (WVF, \cite{platen2007}) and the 
{\it Cosmic Spine} technique \cite{aragonspine2009a} turn to the topological structure of the underlying density field to 
identify void regions and trace the pervasive weblike filamentary network. Both techniques are manifest translations of concepts 
from Morse Theory, which describes the topological structure of a density field on the basis of its singular 
points - minima, maxima and saddle points. 

\begin{figure*}
\vskip -1.0truecm
\begin{center}
  \mbox{\hskip -0.1truecm\includegraphics[width=0.80\textwidth]{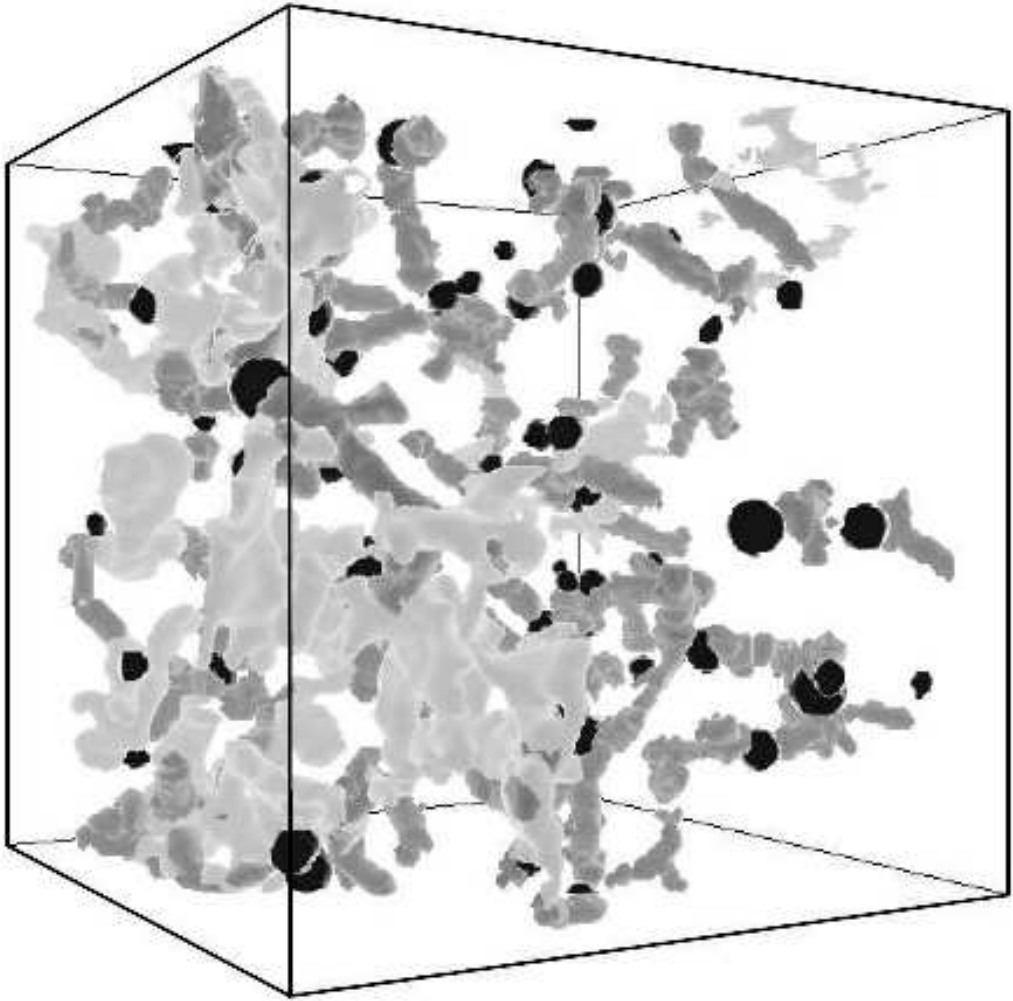}}
\vskip 0.0truecm
\caption{Cosmic web delineated by filaments (dark gray) and walls (light gray). Clusters (dark grey)
are located at the intersection of filaments. Only the largest structures are shown for clarity.}
\label{fig:cosmicwebmmf}
\end{center}
\end{figure*}
\section{the Multiscale Morphology Filter}\ \\
The Multiscale Morphology Filter (MMF), introduced by Arag\'on-Calvo et al. \cite{aragonmmf2007}, is used for the identification and 
characterization of different morphological elements of the large scale matter distribution in the Cosmic Web. 
An example of the morphological structure of the matter distribution of a cosmological N-body simulation can 
be seen in fig.~\ref{fig:cosmicwebmmf} \cite{aragonmmf2007}. The image shows the identified elongated filaments (dark grey), 
sheetlike walls (light grey) and clusters (black dots) in the weblike pattern of the simulation. 

The Multiscale Morphology Filter (MMF) method has been developed on the basis of visualization and feature extraction techniques 
in computer vision and medical research \cite{florack1992}. The technology finds its origin in computer vision research and
has been optimized within the context of feature detections in medical imaging. Frangi et al. \cite{frangi1998} and 
Sato et al. \cite{sato1998} presented its operation for the specific situation of detecting the web of blood vessels in a 
medical image. This defines a notoriously complex pattern of elongated tenuous features whose branching make it closely resemble 
a fractal network. 

The MMF dissects the cosmic web on the basis of the multiscale analysis of the Hessian of the density field. Its 
formulation in terms of its density field {\it scale-space} structure implicitly takes into account the hierarchical nature 
of the matter distribution. Scale Space analysis looks for structures of a mathematically specified type in 
a hierarchical, scale independent, manner.  It is presumed that the specific structural characteristic is quantified by 
some appropriate parameter (e.g.: density, eccentricity, direction, curvature components).  The data is filtered to produce 
a hierarchy of maps having different resolutions, and at each point, the dominant parameter value is selected from the 
hierarchy to construct the scale independent map.  We refer to this scale-filtering processes as a 
\textit{Multiscale morphology filter}.  
 
\subsection{Scale Space}
Crucial for the ability of the method to identify anisotropic features such as filaments and walls is the use of a morphologically 
unbiased and optimized continuous density field retaining all features visible in a discrete galaxy or particle distribution. 
Unless one manages to work directly from the sampled particle or galaxy distribution, possibly via its Voronoi or Delaunay 
tessellation, it is therefore imperative to translate the discrete particle distribution into its DTFE density field,
The morphological intentions of the MMF method render DTFE a key element for translating the particle or galaxy distribution into 
a representative continuous density field $f_{\tiny{\textrm{DTFE}}}$. 

Following the determination of the DTFE density field $f_{\tiny{\textrm{DTFE}}}$, it is smoothed over a range of scales by means 
of a hierarchy of spherically symmetric Gaussian filters $W_{\rm G}$ having different widths $R$:
\begin{equation}
f_{\rm S}({\vec x}) =\, \int\,{\rm d}{\vec y}\,f_{\tiny{\textrm{DTFE}}}({\vec y})\,W_{\rm G}({\vec y},{\vec x})
\end{equation}
where $W_{\rm G}$ denotes a Gaussian filter of width $R$: 
\begin{equation}
W_{\rm G}({\vec y},{\vec x})\, = \,{1 \over ({2 \pi} R^2)^{3/2}}\, \exp \left(- {|{\vec y}-{\vec x}|^2 \over 2 R^2}\right)\,.
\label{eq:filter}
\end{equation}
A pass of the smoothing filter attenuates structure on scales smaller than the filter width. The MMF scale-space analysis 
involves a discrete number of $N+1$ levels, ${n=0,\ldots,N}$. The base-scale $R_0$ is taken to be equal to the pixel scale 
of the raw DTFE density map, while the smoothing radii $R_n$ of the other levels is chosen such that essential 
(sub)structures and features in the density field are resolved. We denote the $n^{th}$ level smoothed version of the DTFE 
reconstructed field $f_{\tiny{\textrm{DTFE}}}$ by the symbol $f_n$. The Scale Space itself is constructed by stacking these 
variously smoothed data sets, yielding the family $\Phi$ of smoothed density maps $f_n$: $\Phi\,=\,\bigcup_{levels \; n} f_n$. 
In our implementation \cite{aragonmmf2007}, we found it sufficient to use only $n=5$ levels, although this may be 
easily expanded. 
 
\begin{figure*}
\vskip -1.0truecm
\begin{center}
  \mbox{\hskip -0.1truecm\includegraphics[width=0.90\textwidth]{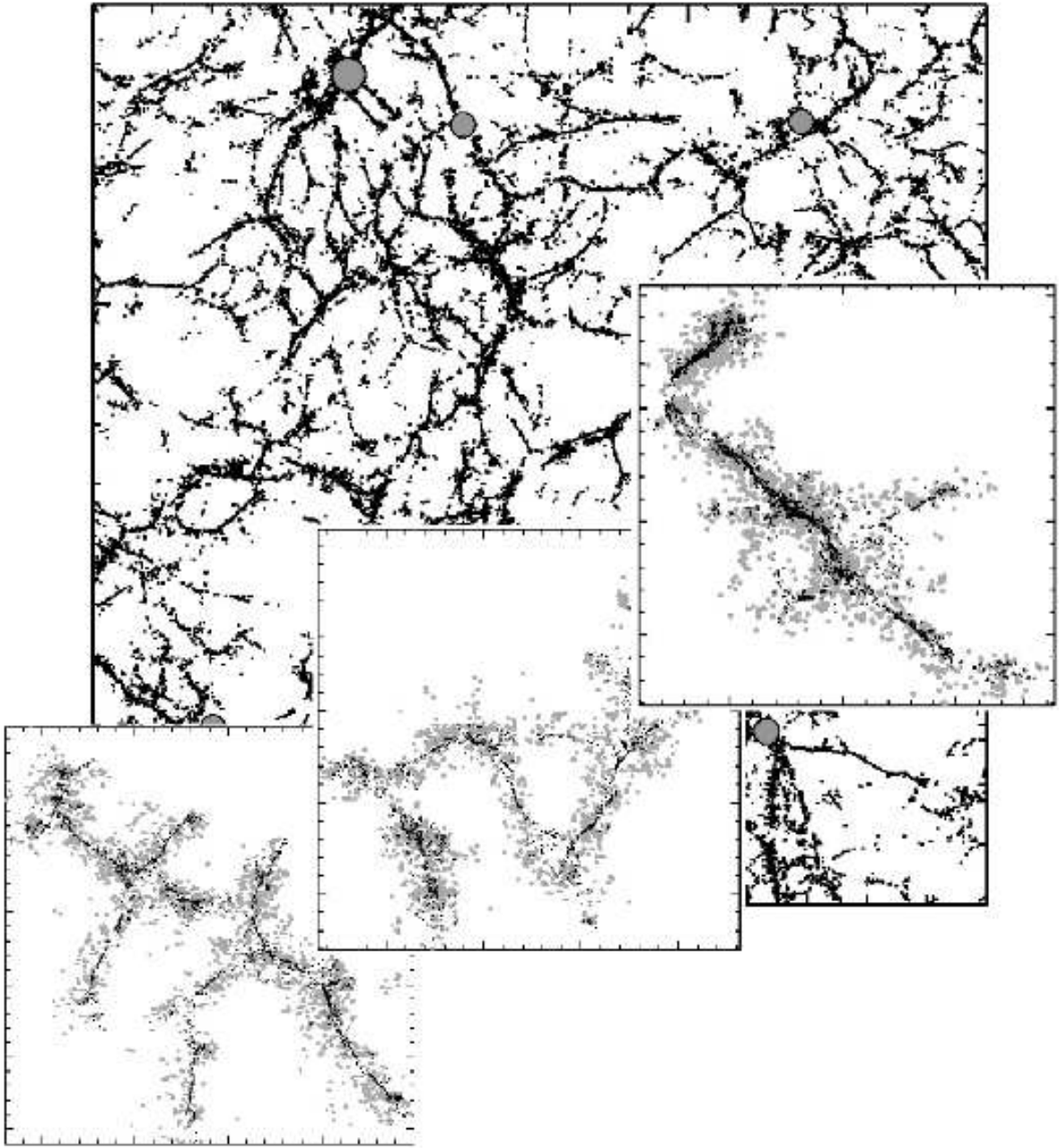}}
\vskip 0.0truecm
\caption{The filamentary network in a $\Lambda$CDM simulation. The filaments were identified 
by means of the MMF technique of Arag\'on-Calvo et al. (2007). The filled (grey) circles correspond 
to clusters with a mass above $10^{14} \hbox{\rm M}_{\odot}$. The inserts contain three specific examples of 
filaments. The gray dots represent the original (simulation) dark matter particles. The spine of 
the filaments (black particles) is the result of the filament compression algorithm of 
Arag\'on-Calvo (2007).}
\label{fig:filamentmmf}
\end{center}
\end{figure*}

\subsection{Density Field Hessian \& Eigenvalues}
A data point can be viewed at any of the scales where scaled data has been generated.  The crux of the concept is that the neighbourhood 
of a given point will look different at each scale. There are potentially many ways of making a comparison of the scale dependence of 
local environment. The MMF employs the Hessian Matrix $\nabla_{ij} f({\bf x})$ of the local density distribution in each of the 
smoothed replicas of the original data, 
\begin{eqnarray}
\frac{\partial^2\,}{\partial x_i \partial x_j}f_S({\vec x})\,=\,f_{\tiny{\textrm{DTFE}}}\,\otimes\,\frac{\partial^2}{\partial x_i \partial x_j} W_{\rm G}(R_{\rm S})\,=\,\hfill\hfill\nonumber \\
\,=\,\int\,{\rm d}{\vec y}\,f({\vec y})\,\frac{(x_i-y_i)(x_j-y_j)-\delta_{ij}R_{\rm S}^2}{R_{\rm S}^4}\,W_{\rm G}({\vec y},{\vec x})\nonumber\\
\end{eqnarray} 
where ${x_1,x_2,x_3}={x,y,z}$ and $\delta_{ij}$ is the Kronecker delta. In other words, the scale space representation of the Hessian 
matrix for each level $n$ is evaluated by means of a convolution with the second derivatives of the Gaussian filter, also known as 
the Marr (or, less appropriately, ``Mexican Hat'') Wavelet. 

At each point of a dataset in the Scale Space view of the data we can quantify the local ``shape" of the density 
field in the neighbourhood of that point by calculating at each point the eigenvalues $\lambda_{a}(\vec{x})$ of 
the Hessian Matrix of the data values, 
\begin{eqnarray}
\qquad \bigg\vert \; \frac{\partial^2 f_n({\vec x})}{\partial x_i \partial x_j}  - \lambda_a({\vec x})\; \delta_{ij} \; 
\bigg\vert  &=& 0,  \quad a = 1,2,3 \nonumber\\
\end{eqnarray}
where the eigenvalues are arranged so that $ \lambda_1 \le \lambda_2 \le \lambda_3 $. The $\lambda_{i}(\vec{x})$ are coordinate independent 
descriptors of the behaviour of the density field in the locality of the point $\vec{x}$ and can be combined to create a variety 
of morphological indicators: they determine the local morphological signal. The corresponding eigenvectors show the local orientation 
of the morphology characteristics. 

\begin{table}
\label{tab:morphmask}
\begin{center}
\begin{tabular} {|c|c|l|}
\hline
\hline
\ &&\\
Structure & $\lambda$ ratios & $\quad$ $\lambda$ constraints \\
\  &&\\
\hline
\ &&\\
Blob     &  $\lambda_1 \simeq \lambda_2 \simeq \lambda_3$ & $\lambda_1 <0\,\,;\,\, \lambda_2 <0 \,\,;\,\, \lambda_3 <0 $  \\
\ &&\\
Line     &  $\lambda_1 \simeq \lambda_2 \ll  \lambda_3$ & $\lambda_1 <0 \,\,;\,\, \lambda_2 <0  $  \\
\ &&\\
Sheet    &  $\lambda_1 \ll    \lambda_2 \simeq \lambda_3$ & $\lambda_1 <0 $    \\
\ &&\\
\hline
\hline
\end{tabular}
\end{center}
\vskip 0.0truecm
\caption{Eigenvalue relationships defining the characteristic morphologies. The
         $\lambda$-conditions describe objects with intensity higher than their
		 local background as clusters, filaments or walls. For voids we would have to reverse the
	     sign of the eigenvalues.}
\vskip -0.5truecm
\end{table}

\subsection{Morphological Identification}
Dependent on the eigenvalue signature and value ratios, we may classify at each scale whether we 
are dealing with a blob, filament or a sheet. Their eigenvalue signatures are listed in table~\ref{tab:morphmask}. 

Following the sequence clusters $\rightarrow$ filaments $\rightarrow$ walls, we identify the regions and scales at which the local 
matter distribution follows the corresponding eigenvalue signature. For each of these morphological components - walls, filaments 
and blobs - the MMF defines a morphology response filter whose form is dictated by the particular morphological feature it seeks to 
extract and whose local value depends on the local shape and spatial coherence of the density field. Detailed expressions for the 
morphology response filter can be found in \cite{aragonmmf2007}. The morphology signal at each location is then defined to be the 
one with the maximum response across the full range of smoothing scales. By means of a density criterion the physically significant 
clumps amongst the blobs are found, while a percolation criterion is invoked to select the physically significant filaments.

Figure \ref{fig:cosmicwebmmf} shows the morphological segmentation of a 150 Mpc $\Lambda$CDM simulation obtained with the MMF. 
The figure illustrates the large scale distribution of matter as an interconnected network of filaments (dark gray) defining the 
boundaries of walls (light gray), and the blobs (black) located at the intersections of the network. 
For clarity fig.~\ref{fig:cosmicwebmmf} only show the largest structures. By restricting the number of structures included it 
is easier to identify the individual components of the Cosmic Web. Each morphological component is well differentiated and occupies, 
by construction, mutually exclusive regions. The ability of the MMF to identify the features of the matter distribution at different 
scales may be inferred from the variety of sizes of blobs. 

Focussing on the filamentary component of the Cosmic Web, fig.~\ref{fig:filamentmmf} shows a slice of the 
simulation box in which MMF identified filaments have been compressed to delineate their spines (see Arag\'on-Calvo \cite{miguelphd2007} 
for a description of the compression algorithm). Gray circles indicate the location of clusters with masses above $10^{14}$ 
M$_{\odot}$ h$^{-1}$. The figure presents the cosmic web as a network of interconnected filaments spread all over the
simulation box. The clusters sit at the intersections or "nodes" of the network \cite{bondweb1996}. The filamentary network 
permeates all regions of space, even the very underdense voids. 

\subsection{Assessment of MMF}
Figures~\ref{fig:cosmicwebmmf} and ~\ref{fig:filamentmmf} testify of the potential of the MMF method for studying the 
structure of the Cosmic Web. Despite its virtues, we may identify various issues. One issue concerns the use of artificial, 
spherically symmetric filters for constructing Scale Space. In the light of the evident anisotropic nature of the cosmic 
matter distribution, this remains a questionable choice. Also, the choice of filter radii is rather artificial, and may 
be improved by finding filter radii more attuned to the particular spectrum of matter flucuations. The singling out of 
individual morphological components prevents a direct and natural connection of these elements in a pervasive 
structure. Finally, in the light of the Cosmic Web theory, with its stress on the key role of tidal field, one may question 
whether an analysis on the basis of the Hessian of the gravitational potential field would not make more sense. 

Partly, these issues are alleviated within the context of the {\it Cosmic Spine} method (see sect.~\ref{sec:cosmicspine}). 

\section{the Watershed Void Finder}
\begin{figure*}
\begin{center}
  \mbox{\hskip -0.1truecm\includegraphics[width=0.95\textwidth]{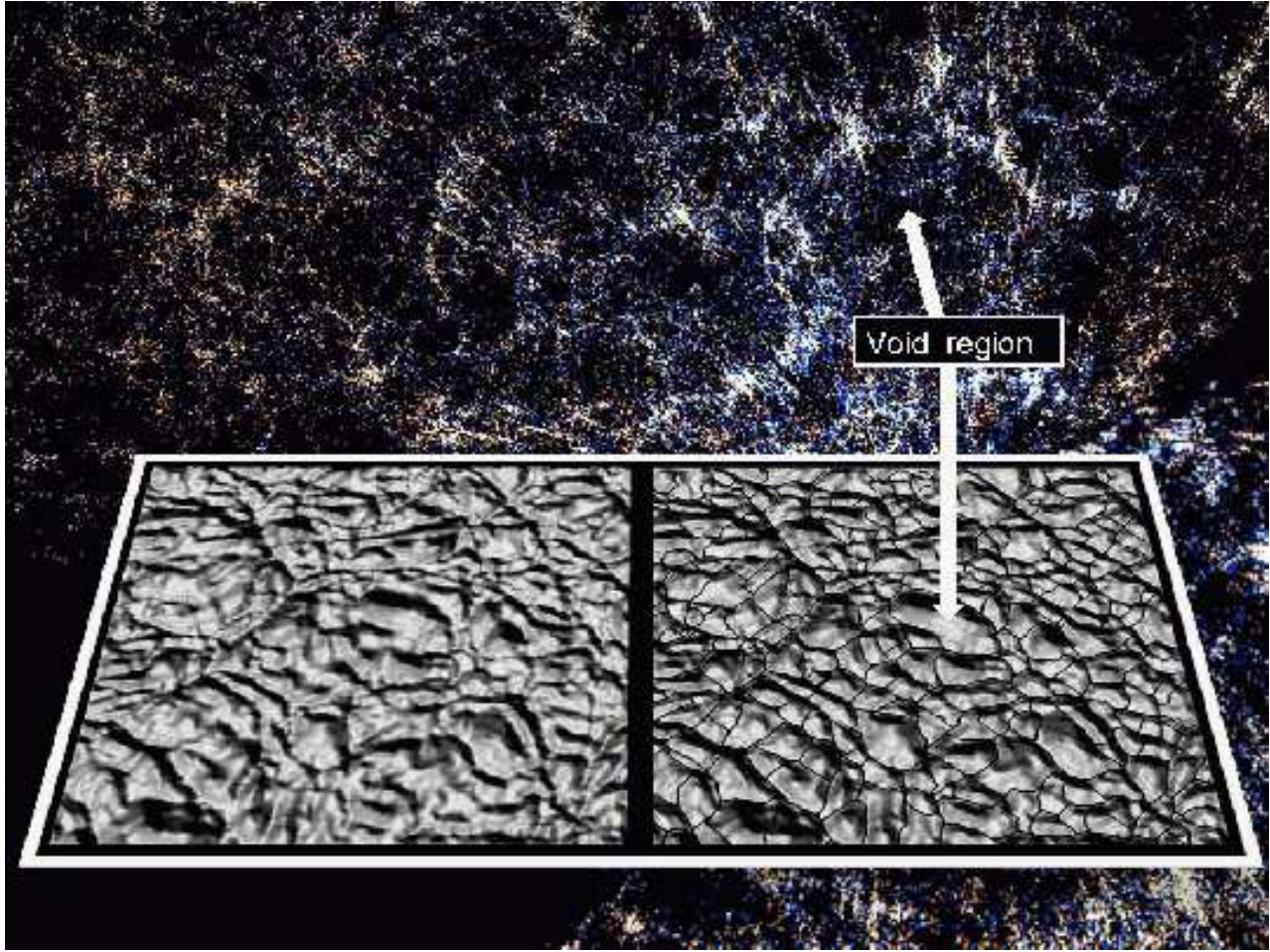}}
\vskip 0.0truecm
\caption{An illustration of a density landscape (left frame) and its segmentation into watershed 
valleys (right frame), surrounded by its watershed lines (solid black). The landscape surfaces are 
depicted against the backdrop of a region selected from the SDSS (DR5) galaxy distribution, 
showing the salient presence of voids.}
\label{fig:wvf}
\end{center}
\end{figure*}

\ \\
\label{sec:wvf}
Voids are enormous regions with sizes in the range of $20-50h^{-1}$ Mpc that are practically devoid of any galaxy and
usually roundish in shape. Forming an essential ingredient of the {\it Cosmic Web} \cite{bondweb1996}, they are surrounded 
by elongated filaments, sheetlike walls and dense compact clusters. Voids are distinctive and striking features of the cosmic web 
(see fig.~\ref{fig:wvf}), yet identifying them and tracing their outline within the complex geometry of the galaxy and 
mass distribution in galaxy surveys and simulations has proven to be a nontrivial issue.

There have been extensive searches for voids in galaxy catalogues and in numerical simulations.  The fact that voids are almost 
empty of galaxies means that the sampling density plays a key role in determining what is or is not a void. There is not an 
unequivocal definition of what a void is and as a result there is considerable disagreement on the precise outline of such a 
region (see e.g. \cite{shandfeld2006}). Moreover, voidfinders are often predicated on building void structures out of cubic 
cells \cite{kauffair1991} or out of spheres (e.g. \cite{hoyvog2002,patiri2006a}). Such methods attempt to synthesize voids from the 
intersection of cubic or spherical elements and do so with varying degrees of success.  Because of the vague and different 
definitions, and the range of different interests in voids, there is a plethora of void identification procedures. The ``voidfinder'' 
algorithm of \cite{elad1997} has been at the basis of most voidfinding methods. However, this succesfull approach will not be able to 
analyze complex spatial configurations in which voids may have arbitrary shapes and contain a range and variety of substructures. 

The watershed-based WVF algorithm of Platen et al. \cite{platen2007} aims to avoid issues of both sampling density and shape. This 
new and objective voidfinding formalism has been specifically designed to dissect in a selfconsistent manner the multiscale 
character of the void network 
and the weblike features marking its boundaries. The WVF is defined with respect to the DTFE density field, assuring optimal sensitivity 
to the morphology of spatial structures and an unbiased probe over the full range of substructure in the mass distribution.

\subsection{Watershed Transform}
\label{sec:watershedtransform}
The {\it Watershed Void Finder} (WVF) is based on the Watershed Transform \cite{beulan1979,meyerbeucher1990,
beumey1993}, a commonly used method in Image Analysis. It is a concept defined within the context of 
mathematical morphology, and was first introduced by Beucher \& Lantuejoul \cite{beulan1979}. It is widely used 
for segmenting images into distinct patches and features. 

The basic idea behind the WST stems from geophysics, where the word {\it watershed} found its 
origin in the analogy of the procedure with that of a landscape being flooded by a rising level 
of water. The WST is used to delineate the boundaries of separate domains, i.e. {\it basins} into 
which yields of e.g. rainfall will collect. Its name may be appreciated from following the analogy 
of a surface in the shape of a landscape. After the surface is pierced at the location of each of the 
minima, and as the water-level rises, a growing fraction of the landscape will be flooded by the water 
in the expanding basins. Ultimately basins will meet at the ridges corresponding to {\it saddle-points} 
in the density field. 

\begin{figure}
\begin{center}
  \mbox{\hskip -0.1truecm\includegraphics[width=0.45\textwidth]{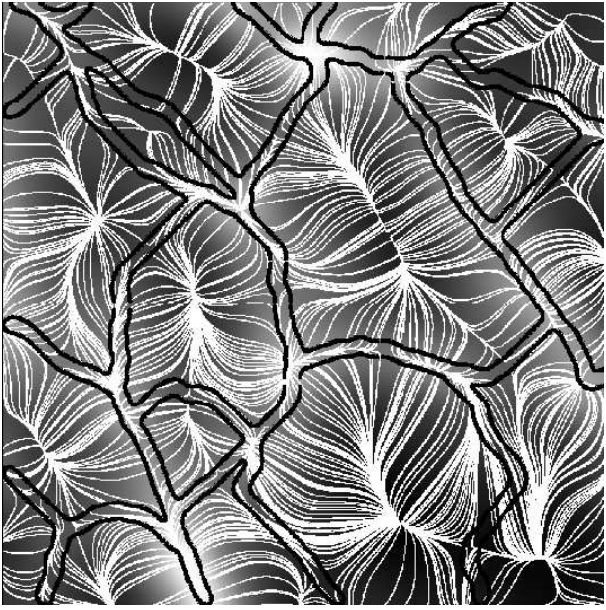}}
\vskip 0.0truecm
\caption{A slice of the density field in a cosmological N-body simulation. Indicated are the 
slope lines (white lines) superimposed on the density field (gray background). The contour of the 
watershed transform is delineated by the the black lines.}
\label{fig:morseflow}
\end{center}
\end{figure}

In many practical situations, including that of the cosmological density fields, we are dealing with a  
landscape of a smooth C$^2$-differentiable density field. The mathematical framework for the topological 
analysis of such surfaces is Morse theory. In the case of a proper Morse function, all critical points 
are non-degenerate with distinct values. Based on the location and nature of the singular points -- minima, 
maxima or saddle points -- and the corresponding \textit{integral lines} or {\it slope lines} 
${\vec s}({\bf x})$ in the density field $f({\bf x})$ (see e.g. fig.~\ref{fig:morseflow}), 
\begin{equation}
{\vec s}({\bf x})\,=\,\nabla f({\bf x})\,,
\end{equation}
one may infer a variety of spatial segmentations (see e.g. \cite{cayley1859,maxwell1870,edelsbrunner2003a,
edelsbrunner2003b,danovaro2003,gyulassy2005}). This may involve regions connected to the maxima, regions 
connected to the minima, or regions connecting a maximum to a particular minimum via the 
\textit{integral lines} of the field. The latter is the Morse-Smale complex of the manifold (also 
see sect.~\ref{sec:cosmicspine} on the Cosmic Spine). The slope lines that define the boundaries of 
adjacent valleys are called \textit{watersheds}, while {\it watershed lines} are the set of slope lines 
emanating from saddle points and connecting to a local maximum or minimum. 

The segmentation of the image is one in regions of influence, defined by a minimum ${\bf y}$ and those 
points ${\bf x}$ whose {\it topographic} distance ${\mathcal T}({\bf x},{\bf y})$ to any of the minima 
in the landscape $f({\bf x})$ is minimal, 
\begin{equation}
{\mathcal T}(x,y)\,\equiv\,{\rm inf} \int_{\Gamma} |\nabla {\mathcal f}(\gamma(s))| ds\,.
\label{eq:topdist}
\end{equation} 
In this definition, the integral denotes the {\it field pathlength} along all paths $\gamma(s)$ in 
the set of all possible paths, $\Gamma$. This concept of  distance is related to the geodesics of 
the surface $f({\bf x})$: the path of steepest descent, which specifies the track a droplet of water 
would follow as it flows down a mountain surface. An illustrative example of a landscape and its 
segmentation into watershed basins surrounded by watershed lines can be seen in fig.~\ref{fig:wvf}.   

\subsection{Cosmological Watersheds}
Extrapolating the application of the watershed transform to other areas of interest, its implementation 
also represents a practical instrument for the segmentation of surfaces and volumes defined by the 
topological structure of cosmological density fields. When studying the topological and morphological structure 
of the cosmic matter distribution in the Cosmic Web, it is convenient to draw the analogy with a landscape (see 
fig.~\ref{fig:wvf}). \textit{Valleys} represent the large underdense voids that define 
the cells of the Cosmic Web. Their boundaries are \textit{sheets} and \textit{ridges}, defining the network 
of walls, filaments and clusters that defines the Cosmic Web. 

The WVF focusses specifically on the valleys, which the Watershed Transform identifies with 
the cosmic voids in the density field. The watershed algorithm holds several advantages with respect 
to other voidfinders:
\begin{itemize}
\item Within an ideal smooth density field (i.e. without
noise) it will identify voids in a parameter free way. No
predefined values have to be introduced. In less ideal, 
and realistic, circumstances a few parameters have to be 
set for filtering out discreteness noise. Their values are 
guided by the properties of the data.
\item The watershed works directly on the topology of the field and 
does not reply on a predefined geometry/shape. By implication the 
identified voids may have any shape. 
\item The watershed naturally places the {\it divide lines} on the
crests of a field. The void boundary will be detected even 
when its boundary is distorted. 
\item The transform naturally produces closed contours. As long 
as minima are well chosen the watershed transform will not be
sensitive to local protrusions between two adjacent voids.
\end{itemize}

\begin{figure*}
\begin{center}
  \mbox{\hskip -0.1truecm\includegraphics[width=0.98\textwidth]{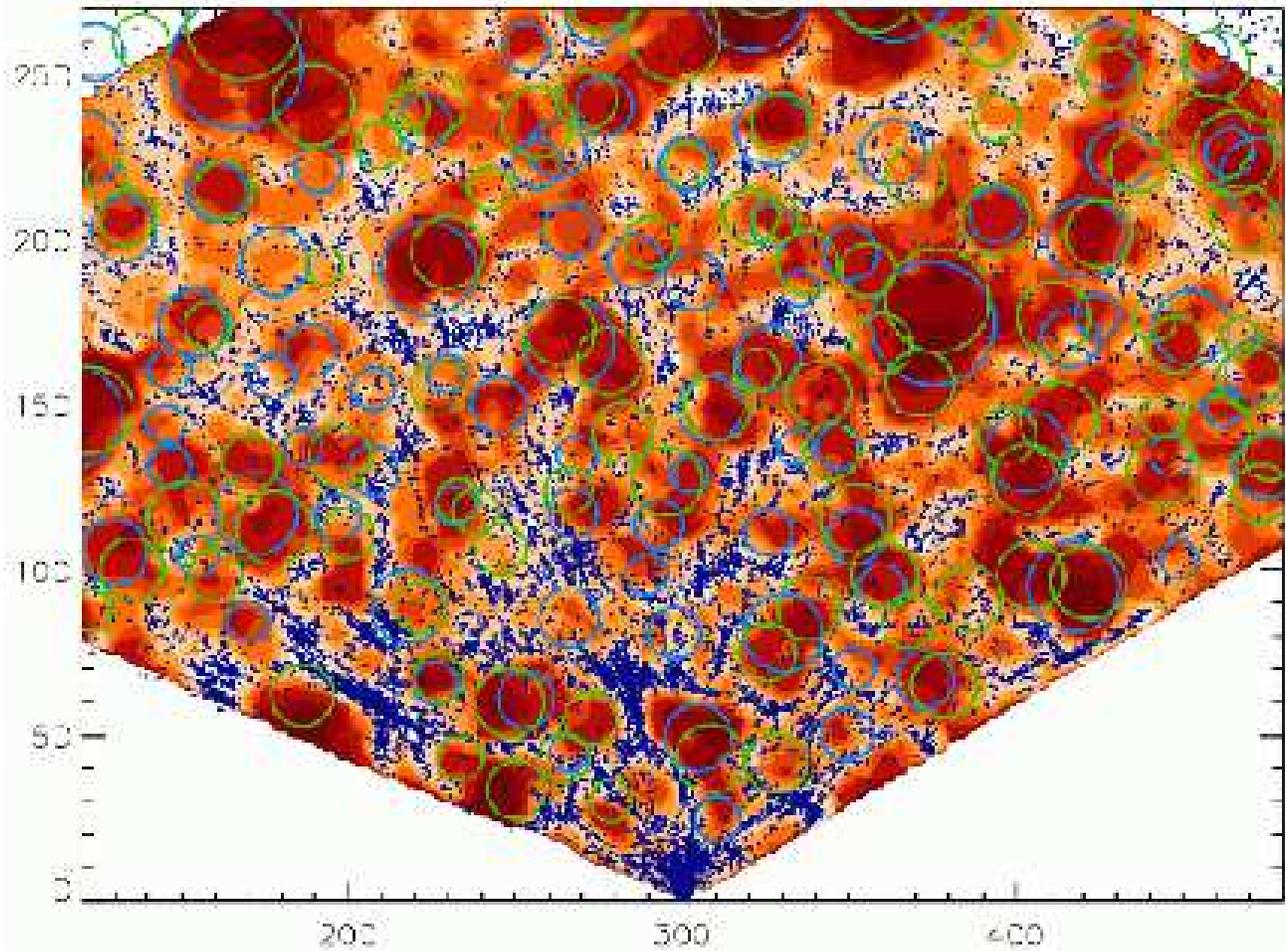}}
\vskip 0.0truecm
\caption{Voids in the SDSS galaxy distribution. The plot shows the distribution of galaxies (SDSS) in a region 
of the SDSS DR7 galaxy redshift survey. Our (ie. Milky Way) central location is at $(300,0)$. The figure depicts 
the distance of galaxies to our Galaxy along radial lines, while the transverse location indicates the direction.  
direction. The light to dark shaded density field is the inferred DTFE density field (dark brown: high density; 
light cream: low density). The blue circles indicate the presence of WVF identified voids. The green circles 
are voids identified with the Voidfinder technique \cite{hoyvog2002}. Image courtesy: Danny Pan.}
\label{fig:sdssvoids}
\end{center}
\end{figure*}

\subsection{Discrete Watershed Transform}
There are many different implementations of the watershed transform and the development of 
such algorithms is still an active field in computer image analysis. An important characteristic 
of the nearly all modern scientific images is their discrete nature. This concerns their 
spatial discreteness, sampled at discrete intervals, and their discrete intensity levels. 
While their discreteness allows the use of highly efficient algorithms, it does also involve 
complications due to the intrinsic difficulty in identifying saddle points from a discrete local 
neighborhood and of accurately extracting slope lines. On the other hand, by removing faint 
features the discretization also helps to remove some artefacts of the watershed transform without 
the need of pre or postprocessing.

Several methods have been devised in an attempt to alleviate the limitations imposed by discrete 
images for the extraction of critical points. Among them, the \textit{discrete watershed transform} 
algorithm \cite{beucher1982} represents a simple and elegant way of identifying the watershed separatrices 
and it can be shown to converge to the continuous case. The watershed transform emulates the flooding of 
valleys. As these are flooded the newly formed ``lakes" increase their area until they touch each other. 
The points where two or more lakes converge are marked and the algorithm continues until all the pixels 
in the image have been flooded. At the end of the flooding process the image will be \textit{segmented} 
into individual regions or catchment basins sharing a local minima. The resulting watershed transform is 
defined by the set of all points marked as the dividing boundaries between two or more basins. 

\subsection{the WVF algorithm}
We have implemented a version of the discrete watershed algorithm of Beucher \cite{beucher1982}, 
and incorporated it in the WVF void identification procedure. The WVF consists of the 
following steps: 
\begin{itemize}
\item {\bf DTFE}: Given a point distribution (N-body, redshift
survey), the Delaunay Tessellation Field Estimator is used to define 
a continuous density field throughout the sample volume. 
\item {\bf Grid Sampling}: 
For practical processing purposes the DTFE field is sampled 
on a grid. The optimal grid size has to assure the resolution of  
all morphological structures while minimizing the number 
of needed gridcells.  
\item {\bf Rank-Ordered Filtering}: The DTFE density field is
adaptively smoothed by means of {\it Natural Neighbour Maxmin and
Median} filtering. This involves the computation of the median,
minimum or maximum of densities within the {\it contiguous Voronoi
cell}, the region defined by a point and its {\it natural neighbours
}.
\item {\bf Contour Levels}: The image is transformed into a discrete
set of density levels. The levels are defined by a uniform
partitioning of the cumulative density distribution.
\item {\bf Pixel Noise}: With a morphological {\it opening} and 
{\it closing} of 2 pixel radius we further reduce pixel by pixel fluctuations.
\item {\bf Field Minima}: The minima in the smoothed density field are
identified as the pixels (grid cells) which are exclusively surrounded 
by neighbouring grid-cells with a higher density value.
\item {\bf Flooding}: The {\it flooding procedure} starts at the
location of the minima. At successively increasing flood levels the
surrounding region with a density lower than the corresponding
density threshold is added to the {\it basin} of a particular
minimum.
\item {\bf Segmentation}: Once a pixel is reached by two distinct
basins it is identified as belonging to their segmentation
boundary. By continuing this procedure up to the maximum density
level the whole region is segmented into distinct {\it void
patches}.
\item {\bf Hierarchy Correction}: A correction is
necessary to deal with effects related to the intrinsic hierarchical
nature of the void distribution. The correction involves the removal
of segmentation boundaries whose density is lower than some
density threshold.
\end{itemize}

\subsection{WVF by example: Voids in SDSS}
A telling impression of the practical implementation of the watershed voidfinding method may be 
obtained from fig.~\ref{fig:sdssvoids}. The watershed procedure has been applied to the
SDSS DR7 galaxy survey. 

The Sloan Digital Sky Survey is the largest campaign for mapping the spatial distribution of galaxies in 
the local universe. It has imaged 25\% of the sky: with the latest DR7 data release it completed 
the survey of objects over 8423 sq. deg. area on the sky. Around 230 million objects were imaged in 
5 photometric bands, out to apparent magnitude $\sim 25.1$ (r band, $\lambda \sim 628$ nm)\footnote{In astronomy, the 
brightness of objects is quantified in terms of magnitude. Originally defined by Hipparchus, and in 
accordance with the logarithmic sensitivity of the eye, it concerns the logarithm of the brightness $I$ of the objects, 
$M \propto -2.5 \log{I}$. Note that a lower magnitude implies a brighter object. The human eye can see 
stars brighter than $m=6$. The star Vega defined the zeropoint, $m=0$, while Sirius, the brightest star on 
the sky (except for the Sun) has $m=-1.47$. A galaxy with a magnitude $m=17.7$ is really very faint !}. 
Of 928,567 galaxies, those brighter than $r \sim 17.7$, the redshift (distance) was measured. 

Fig.~\ref{fig:sdssvoids} shows the DTFE density field inferred from the discrete galaxy distribution 
in a wide slice through the DR7 sample. It concerns a region which goes out to a redshift $z\sim 0.107$, 
ie. out to a distance of $d \sim 430$ Mpc. The galaxies are superimposed as blue dots. Note the lower 
spatial resolution of the survey at larger distances: as their distance increases only the brightest 
galaxies can be seen, resulting in a reduced spatial sampling. 

The blue circles indicate the position and size (but not shape, and outline) of the voids in 
the SDSS. WVF identified $\sim 1000$ voids in the SDSS. The WVF voids are compared with those 
found by the voidfinder of Hoyle \& Vogeley \cite{hoyvog2002}, indicated as green circles. The location of the 
larger voids match quite well, although there are differences concerning the size of the 
identified voids. A more substantial mismatch exists for smaller voids. 

\subsection{Assessment of WVF}
With respect to the other void finders the watershed algorithm has several advantages. Because it identifies a 
void segment on the basis of the crests in a density field surrounding a density minimum it is able to trace 
the void boundary even though it has a distorted and twisted shape. Also, because the contours around well chosen 
minima are by definition closed the transform is not sensitive to local protrusions between two adjacent voids. 
The main advantage of the WVF is that for an ideally smoothed density field it is able to find voids in an entirely 
parameter free fashion. 

\section{the Cosmic Spine \& Morse Theory}
\label{sec:cosmicspine}
The \textit{SpineWeb} method \cite{aragonspine2009a} addresses the topological structure of the landscape, defined by its 
singularities and their mutual connections, that of minima, maxima and saddle points. It invokes the local adjacency 
properties of the boundaries between the watershed basins to trace the critical points in the density field and the 
separatrices defined by them. The separatrices are classified into \textit{walls} and the {\it spine}, the network of 
filaments and nodes in the matter distribution. The SpineWeb formalism is intimately related to Morse theory, in which 
it finds a solid mathematical foundation. 

In the discussion of the WVF (sec.~\ref{sec:wvf}), we have seen that the watershed transform is a key instrument 
for the segmentation of a density field, and as such is also ideally suited for tracing the boundaries between 
the identified segments. With this in mind we have invoked the watershed transform in a more extensive 
framework for studying both identity and topology of the cosmic web and its various constituents. The result is 
the \textit{SpineWeb} method, a complete framework for the identification of voids, walls and filaments. 
Based on the {\it watershed segmentation} of the cosmic density field, the method invokes the local properties of 
the regions adjacent to the critical points, which define its {\it separatrices}. An inspection of the {\it flow lines} 
in a 2-D density field, and the corresponding separatrices, forms a telling illustration of the idea 
(fig.~\ref{fig:morseflow}).  

\subsection{SpineWeb and the Cosmic Web}
The \text{SpineWeb} technique is based on the following key aspects of the topological structure of the Cosmic Web: 
\begin{itemize}
\item[$\bullet$] The Cosmic Web is an interconnected system of walls, filaments and clusters. High resolution N-body 
simulations indicate that all the elements of the Cosmic Web are interconnected. As mass resolution increases, 
one finds structures that form the connection between otherwise apparently isolated objects.
\item[$\bullet$] Walls are the defining boundaries between adjacent voids. The analogy between the 
watershed transform defining the boundary between underdense basins and walls defining the boundaries 
between voids is one of the major justifications of the method presented here. 
\item[$\bullet$] Filaments run between clusters and across the intersection of walls. {\it Membranes} permeate 
the space between adjacent filaments. 
\end{itemize}

\bigskip
An important aspect of the SpineWeb formalism is the practical role of the watershed transform in computing the 
Morse complex of a density field. 
In sect.~\ref{sec:wvf} we have shortly described how the {\it flow field} of a landscape can be used towards 
defining a variety of spatial segmentations (see e.g. \cite{cayley1859,maxwell1870,edelsbrunner2003a,
edelsbrunner2003b,danovaro2003,gyulassy2004,gyulassy2005,gyulassy2007}). These are based on the connection between the singularities 
- maxima, minima and sadddle points - in the landscape by means of the \textit{slope lines}. Within 
this context, the {\it watershed lines} are the set of slope lines emanating from saddle points and connecting 
to a local maximum or minimum, with the ones connecting to maxima defining the watershed boundaries around the 
\textit{valleys}. 

The Morse theoretical basis of the SpineWeb formalism relates it to the {\it skeleton analysis} of the 
Cosmic Web \cite{novikov2006,sousbie2008a,sousbie2008b}, which is also based on Morse theory \cite{colombi2000,pogosyan2009}. 
It formed the basis for the development of an elegant and mathematically rigorous tool for filament identification, which 
in the meantime has also been extended towards tracing a range of morphological features \cite{sousbie2009}. Its present 
implementation refers to only one specific scale. The SpineWeb procedure is being generalized to a fully 
multiscale formalsim. 

\subsection{SpineWeb procedure}
The SpineWeb procedure is based on an implementation of the discrete watershed transform code 
\cite{beucher1982,roerdink2000}. The code starts by identifying and labeling all the local minima 
present in the density field as the pixels which have the lowest value among its 26 neighbours. Subsequently, for 
each voxel the neighbour with lowest density is determined. By pursuing this, the procedure iteratively traces 
the maximum gradient paths until they converge to their local minima. This topographical distance 
algorithm yields a fast segmentation of space into locally connected underdense regions. 

In the second step we identify the voxels embedded in the watershed transform. For this, a 
\textit{local immersion algorithm} is used. All the voxels located at the boundaries between two or more 
regions are identified, Subsequently, this subset of voxels is sorted in density, followed by the 
application of the standard immersion algorithm. The advantage of this two-step process is that there 
is no need for sorting the complete density field. Instead, only the voxels in the boundaries around the 
watershed segments are addressed. 

\begin{figure}
  \centering
  \includegraphics[width=0.5\textwidth]{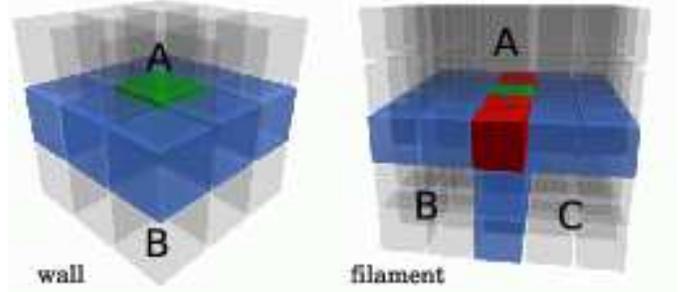}
    \caption{Local neighborhood around a voxel (green) inside a wall (left) and filament (right).            
        The blue voxels indicate walls and the red voxels filaments. The light gray cubes here represent
	voxels inside voids. The voxel inside a wall has two adjacent voids inside its neighborhood (A and B) while 
        the voxel inside a filament has three adjacent voids (A, B and C). From Arag\'on-Calvo et al. 2009a.}
  \label{fig:simple_neighbourhoods}
\end{figure} 

\begin{figure*}
\begin{center}
  \mbox{\hskip -0.1truecm\includegraphics[width=0.98\textwidth]{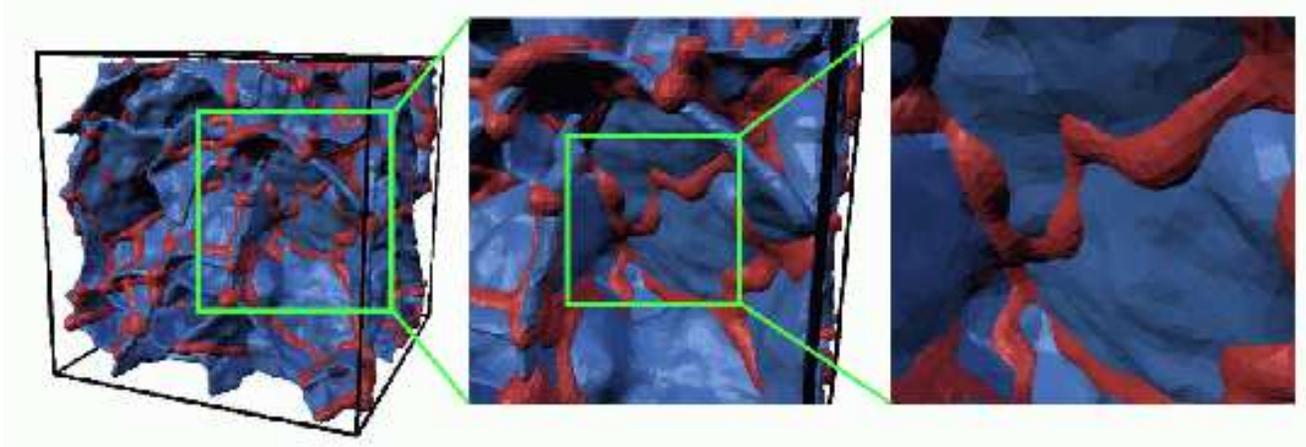}}
\vskip 0.0truecm
\caption{Surfaces enclosing the voxels which are identified as belonging to walls (blue) and 
        filaments (red) within a cubic region of $50 h^{-1}$ Mpc. The frames show how both morphological 
        components are connected and intertwined, a nice illustration of the intimate relationship between 
        filaments and walls. For visualization purposes the surfaces are smoothed with a Gaussian kernel of 
        $\sigma=2$ voxels. The zoom-ins onto the cosmic spine in a subregion of the $50 h^{-1}$ Mpc (central 
        and righthand frame), highlight these intricate connections between wall surfaces, filamentary edges 
        and cluster nodes. From Arag\'on-Calvo et al. 2009a.}
\label{fig:simspine}
\end{center}
\end{figure*}

In the final step, each boundary voxel is assigned a morphological label as \textit{void}, \textit{wall} or 
\textit{filament} on the basis of their local neighbourhood, ie. the 26 adjacent voxels to each boundary 
element. The assigned morphology follows a set of simple but effective criteria based on the number of adjacent 
voids/basins $\mathcal{N}_{\textrm{\tiny{voids}}}$ to a given boundary voxel (see fig.~\ref{fig:simple_neighbourhoods}), 
\begin{equation}
\mathcal{N}_{\textrm{\tiny{voids}}} \left\{
 \begin{array}{rl}
     \qquad   =  \;\;\; 1, \qquad & \text{void} \\
     \qquad   =  \;\;\; 2, \qquad & \text{wall} \\
     \qquad \geq \;\;\; 2, \qquad & \text{filament}
 \end{array} \right.
\label{eq:neighbour_morphology}
\end{equation}
\noindent It is important to emphasize that these criteria for identifying filaments and walls from the 
watershed transform relies \textit{only} on the topology of the density field. 

\subsection{Filaments, Sheets and the Spine}
The identification of walls with locations marked by two adjacent voids is straightforward: they are sheets 
that collect matter from their surrounding voids. The identification of filaments depends strongly 
on the resolved cellular structure of the Cosmic Web. Filaments that are related to voids whose 
outline is visible in the density map are succesfully traced. However, when the density field representation 
is too coarse, or when the substructure of voids is insufficiently resolved, the procedure tends to 
produce a variety of less pronounced and smaller filamentary features that have not been 
identified as spinal filaments. In the hierarchical multiscale extension of the SpineWeb method 
this is directly addressed.

Figure~\ref{fig:simspine} shows the full 3D network of filaments (red) and walls (blue) in a 
cosmological N-body simulation. It concerns a $\Lambda$CDM universe inside a box of 200 $h^{-1}$ Mpc, 
restricted to the dark matter particles. Initial conditions were generated on a $512^3$ grid with 
$\Omega_m = 0.3$, $\Omega_{\Lambda} = 0.7$, $\sigma_8 = 0.9$ and $h = 0.73$. Lower-resolution versions 
of $256^3$, $128^3$ and $64^3$ particles were generated from the same initial conditions. The depicted 
spine structure focuses on the largest filaments and walls in the particle distribution, following 
from the analysis of the low-resolution $64^3$ dataset. This lower resolution corresponds to a cut-off 
scale of roughly $3$ h$^{-1}$Mpc. 

For visualization purposes, the walls and filaments are separately smoothed with a Gaussian kernel of $\sigma=2$ voxels.
Figure~\ref{fig:simspine} clearly shows the three-dimensional nature of the filament-wall network and the close relation 
between both: filaments define an interconnected web and walls ``fill" the spaces in between filaments forming 
a watertight network of voids. Walls can be seen as extended membranes with somewhat irregular surfaces on small scales. 
On larger scales walls can be considered semi-planar structures, although it is possible to find examples of walls with more 
convoluted shapes reflecting equally complex local environments. Filaments (here an individual filament is defined 
as the filament between two intersections points) also present small scale features but overall they can 
be considered semi-linear, even though one can find very twisted structures. Short filaments appear to be 
more straight than long ones. 

\begin{figure*}
\vskip -1.0truecm
\begin{center}
  \mbox{\hskip -0.1truecm\includegraphics[width=0.60\textwidth]{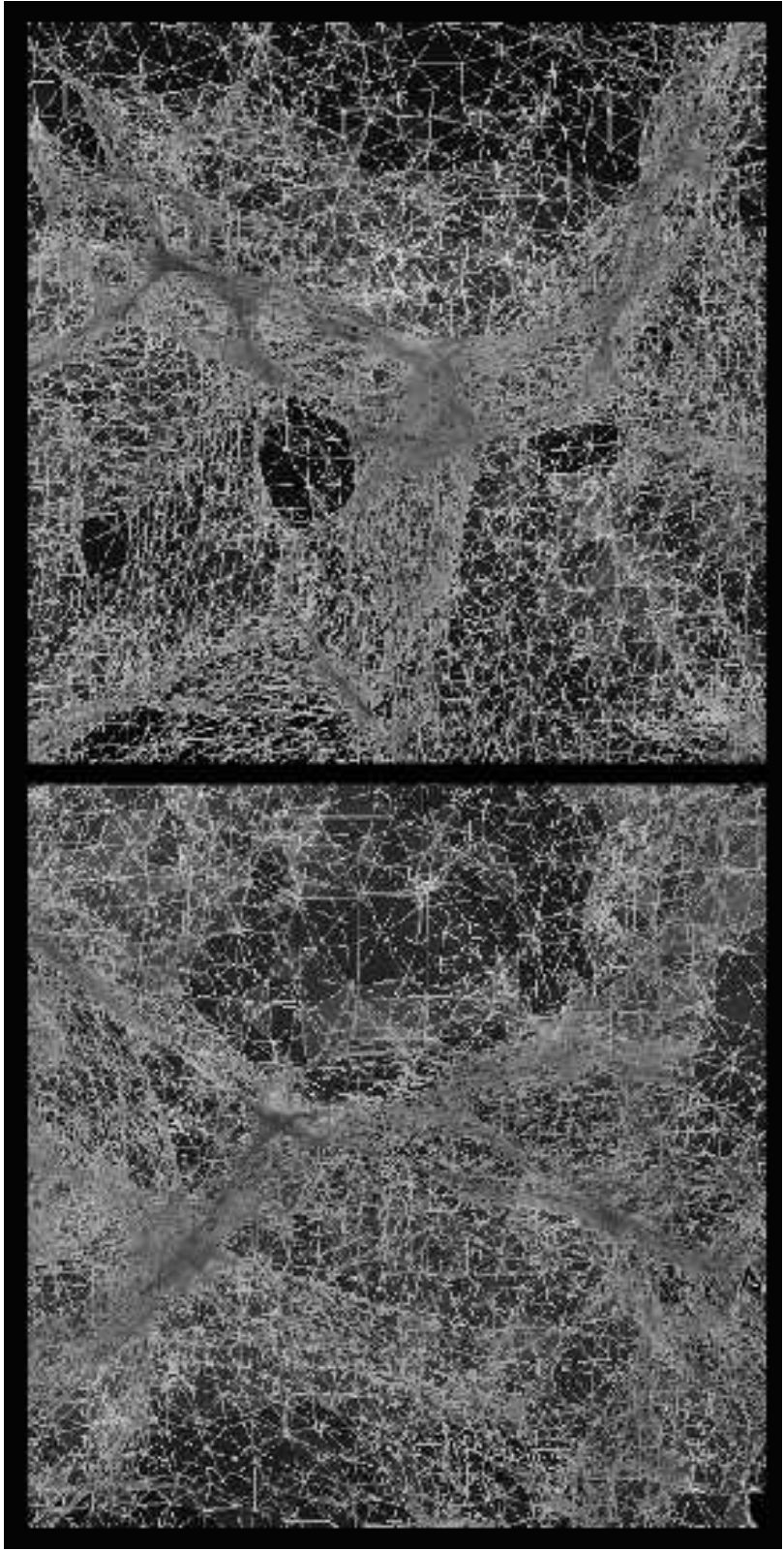}}
\vskip 0.0truecm
\caption{Illustration of the tessellation-based SpineWeb procedure. Both panels show the cosmic 
spine of the same $\Lambda$CDM simulation, seen from different viewpoints. They show the points 
and their mutual Delaunay (edge) connections (yellow lines) which belong to the filaments and 
sheets in the Cosmic Web. The morphological identity of each point is determined on the basis 
of a density comparison with their natural neighbours.}
\label{fig:tessspine}
\end{center}
\end{figure*}

\begin{figure*}
\begin{center}
  \mbox{\hskip -0.1truecm\includegraphics[width=0.99\textwidth]{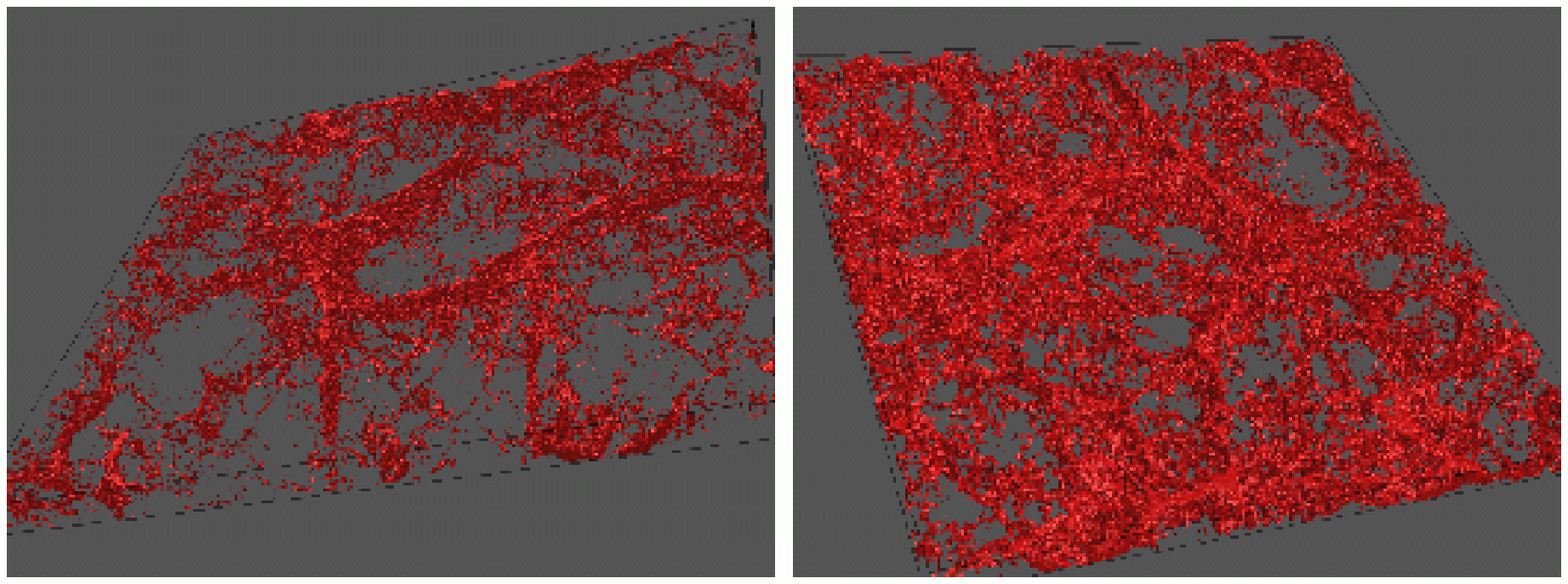}}
\vskip 0.0truecm
\caption{Examples of {\it alpha shapes} of the LCDM GIF simulation. Shown are central slices through 
two alpha shapes (top: low alpha; bottom: high alpha). The image shows the sensitivity of alpha shapes 
to the topology of the matter distribution. Image courtesy of Bob Eldering. From: Vegter et al. 2009.}
\label{fig:alphashape}
\end{center}
\end{figure*}

\subsection{the Hierarchical Spine}
An important aspect of our SpineWeb method is that it is an intrinsically scale-free 
method, starting from a scale-free reconstruction of the density field. The DTFE density field 
guarantees an optimal and unbiased representation of the hierarchical nature and anisotropic morphology 
of cosmic structure. Having guaranteed the capability of invoking a full scale-free {\it Scale-space} 
representation of cosmic structure, the watershed procedure not only traces the outline of filaments and 
sheets, but may also be extended towards doing so over a range of scales in order to address their 
hierarchical structure.  

At present, the spine formalism has indeed been extended to a genuine multiscale formalism. This 
allows the identification of the spine over a range of scales. The spine segments on a range of scales are 
subsequently related to each other, allowing the construction of hierarchical filament, wall and 
void trees (see upcoming publication \cite{aragonspine2009b}). 

The hierarchical spine procedure has demonstrated that seemingly unrelated substructures at one 
scale, do actually show up in the spine of higher resolution realizations of the density field. This 
is a direct reflection of the presence of the cosmic web's topological outline in the primordial 
density field, and assures the fact that filaments always lie at the intersection of walls and 
form the edges of empty cells. 

\subsection{Tessellation Spines}
While in the present implementation we determine the spine via the DTFE density field sampled 
on a regular grid, a highly promising innovation will be to circumvent the density field 
reconstruction and gridbased spineweb identification procedure entirely by working directly 
on the Delaunay tessellation defined by the point sample itself. 

From the DTFE discussion in sect.~\ref{sec:dtfefund} we have learnt that the Delaunay/Voronoi 
tessellation contains all necessary information on the local density and geometry in the sample. 
Instead of first determining the density first, we could define local criteria for the Delaunay 
tetrahedra for tracing the singularities and the separatrices in the corresponding field. It will 
render the Spineweb procedure independent of an artificial gridscale, grid geometry and save the 
consumption of CPU time for computing the DTFE field. The extraction of the complete hierarchy of 
weblike structures could be accomplished in one calculation, that of the tessellation.

Various visualization and image processing studies have been seeking to define Morse theory 
and watershed segmentation algorithms that work directly via the Delaunay tessellation of 
the discrete dataset \cite{danovaro2003,edelsbrunner2003b,bremer2004,magillo2007,danovaro2007,gyulassy2007}. 

The high potential of this strategy is illustrated by fig.~\ref{fig:tessspine}. It involves the 
a first test implementation of the SpineWeb procedure directly on the Delaunay grid 
\cite{aragonspine2009c}. Instead of resorting to the 26 neighbours on a regular grid, the 
morphological evaluations (eqn.\ref{eq:neighbour_morphology}) for each point in the sample 
concerns its {\it natural neighbours}, the points with whom it shares a Delaunay tetrahedron. 
The two panels of fig.~\ref{fig:tessspine} shows the points in the filaments and sheets of 
the Cosmic Web, along with their Delaunay (edge) connections. The intricate weblike features 
are clearly represented in unprecedented detail, both the rather tenuous thin membranes 
and the dense and abundantly branching filaments. 

\section{Topology \& Alphashapes}
\label{sec:alphashape}
A final tessellation-based technique for the analysis of aspects of the Cosmic Web relates to its 
topological aspects. {\it Alpha shape} is a description of the (intuitive) notion of the shape of a discrete point set. 
{\it Alpha-shapes} of a discrete 
point distribution are subsets of a Delaunay triangulation and were introduced by Edelsbrunner and collaborators 
\cite{edelsbrunner1983,mueckephd1993,edelsbrunner1994,edelsbrunner2002}. Alpha shapes are generalizations of the convex 
hull of a point set and are concrete geometric objects which are uniquely defined for a particular point set. Reflecting 
the topological structure of a point distribution, it is one of the most essential concepts in the field of Computational 
Topology \cite{dey1998,vegter2009,zomorodian2005}. 

If we have a point set $S$ and its corresponding Delaunay triangulation, we may identify all {\it Delaunay simplices} 
-- tetrahedra, triangles, edges, vertices -- of the triangulation. For a given non-negative value of $\alpha$, the 
{\it alpha complex} of a point sets consists of all simplices in the Delaunay triangulation which have 
an empty circumsphere with squared radius less than or equal to $\alpha$, $R^2\leq \alpha$. Here ``empty'' means that the 
open sphere does not include any points of $S$. For an extreme value $\alpha=0$ the alpha complex merely consists of 
the vertices of the point set. The set also defines a maximum value $\alpha_{\rm max}$, such that for $\alpha \geq 
\alpha_{\rm max}$ the alpha shape is the convex hull of the point set. 

The {\it alpha shape} is the union of all simplices of the alpha complex. Note that it implies that although the alpha shape 
is defined for all $0\leq \alpha < \infty$ there are only a finited number of different alpha shapes for any one point set. 
The alpha shape is a polytope in a fairly general sense, it can be concave and even disconnected. Its components can be 
three-dimensional patches of tetrahedra, two-dimensional ones of triangles, one-dimensional strings of edges and 
even single points. The set of all real numbers $\alpha$ leads to a family of shapes capturing the intuitive notion of 
the overall versus fine shape of a point set. Starting from the convex hull of a point set and gradually decreasing $\alpha$ the shape 
of the point set gradually shrinks and starts to develop cavities. These cavities may join to form tunnels and 
voids. For sufficiently small $\alpha$ the alpha shape is empty. 

Following this description, one may find that alpha shapes are intimately related to the topology of a point set. 
As a result they form a direct and unique way of characterizing the topology of a point distribution. A complete 
quantitative description of the topology is that in terms of Betti numbers $\beta_{\rm k}$ and these may indeed 
be directly inferred from the alpha shape. The first Betti number $\beta_0$ specifies the number of 
independent components of an object. In this context $\beta_1$ may be interpreted as the number of independent 
tunnels, and $\beta_2$ as the number of independent enclose voids. The $k^{th}$ Betti number effectively counts 
the number of independent $k$-dimensional holes in the simplicial complex. 

Applications of alpha shapes have as yet focussed on biological systems. Their use in characterizing the topology 
and structure of macromolecules. The work by Liang and collaborators \cite{edelsbrunner1998,liang1998a,liang1998b,liang1998c}   
uses alpha shapes and betti numbers to assess the voids and pockets in an effort to classify complex protein structures, 
a highly challenging task given the 10,000-30,000 protein families involving 1,000-4,000 complicated folds. Given the interest 
in the topology of the cosmic mass distribution \cite{gott1986,mecke1994,schmalzing1999}, it is evident that {\it alpha shapes} 
also provide a highly interesting tool for studying the topology of the galaxy distribution and N-body simulations of cosmic 
structure formation. Directly connected to the topology of the point distribution itself it would discard the need of 
user-defined filter kernels. 

In a recent study Vegter et al. \cite{vegter2009} computed the alpha shapes for a set of GIF simulations of cosmic structure 
formation (see fig.~\ref{fig:alphashape}). On the basis of a calibration of the inferred Minkowski functionals and 
Betti numbers from a range of Voronoi clustering models their study intends to refine the knowledge of 
the topological structure of the Cosmic Web. 

\section*{Acknowledgement}
We are very grateful to Willem Schaap for his contributions to substantial parts of the work presented 
in this review. We also thank Gert Vegter for very useful discussions on various aspects of the presented 
work, and Danny Pan for permission to use fig.~\ref{fig:sdssvoids}. Finally, we wish to express our gratitude to 
the organizers of the ISVD09 conference, foremost F. Anton, for the invitation to present this review. 



%
\bibliographystyle{IEEEtran}
\bibliography{IEEEabrv,isvd09wey}

\end{document}